\documentclass{egpubl}
\usepackage{pg2020}
 
\SpecialIssuePaper         

\usepackage[T1]{fontenc}
\usepackage{dfadobe}  

\usepackage{cite}  
\BibtexOrBiblatex
\electronicVersion
\PrintedOrElectronic
\ifpdf \usepackage[pdftex]{graphicx} \pdfcompresslevel=9
\else \usepackage[dvips]{graphicx} \fi

\usepackage{egweblnk} 

\usepackage{url}
\usepackage{color}
\usepackage{lipsum}
\usepackage{graphicx}
\usepackage[table,xcdraw]{xcolor}
\usepackage{eurosym}
\usepackage{amsfonts}

\newcommand\XX[1]{{\color{black}{#1}}} 

\title[\textit{Slice and Dice}]%
      {\textit{Slice and Dice}: \\A Physicalization Workflow for Anatomical Edutainment\vspace{-12pt}}

\author[R.G.Raidou, M.E.Gr{\"o}ller \& H.-Y.Wu]
       {Renata G. Raidou,$\!^{1,2}\;\;$%
       M. Eduard Gr{\"o}ller,$\!^{2,3}\;\;$%
       Hsiang-Yun Wu$\,^{2}\;\;$%
       \\ \vspace{-20pt}
        \footnotesize $^1$\,University of Groningen, the Netherlands, $^2$\,TU Wien, Austria, $^3$\,VRVis Research Center, Austria\\
       }

\begin{document}

\maketitle
\begin{abstract}
During the last decades, anatomy has become an interesting topic in education---even for laymen or schoolchildren.
As medical imaging techniques become increasingly sophisticated, virtual anatomical education applications have emerged. 
Still, anatomical models are often preferred, as they facilitate 3D localization of anatomical structures.
Recently, data physicalizations (i.e., physical visualizations) have proven to be effective and engaging---sometimes, even more than their virtual counterparts.
So far, medical data physicalizations involve mainly 3D printing, which is still expensive and cumbersome. 
We investigate alternative forms of physicalizations, which use readily available technologies (home printers) and inexpensive materials (paper or semi-transparent films) to generate crafts for anatomical edutainment.
To the best of our knowledge, this is the first computer-generated crafting approach within an anatomical edutainment context.
Our approach follows a cost-effective, simple, and easy-to-employ workflow, resulting in assemblable data sculptures (i.e., semi-transparent sliceforms).
It primarily supports volumetric data (such as CT or MRI), but mesh data can also be imported. 
An octree slices the imported volume and an optimization step simplifies the slice configuration, proposing the optimal order for easy assembly. 
A packing algorithm places the resulting slices with their labels, annotations, and assembly instructions on a paper or transparent film of user-selected size, to be printed, assembled into a sliceform, and explored.
We conducted \XX{two user studies to assess our approach, demonstrating that it is} an initial \XX{positive} step towards the successful creation of interactive and engaging anatomical physicalizations.

\ccsdesc[500]{Human-centered computing~Visualization application domains}
\ccsdesc[500]{Applied computing~Life and medical sciences}

\printccsdesc   
\end{abstract}

\section{Introduction}
\label{sec:intro}

Patient education and involvement in treatment have become standard practice in many places around the world~\cite{gruman2010patient}. 
At the same time, there is increasing public interest in learning anatomy and physiology. 
For laymen without medical training or for a schoolchild, understanding the available information and recognizing structures within medical images, such as CT or MRI, would be overwhelming. 
Hence, anatomical illustrations (2D or 3D, in books or on-screen, static or interactive) are often employed. 
Sometimes anatomical models are preferred, as they facilitate 3D localization of structures within the human anatomy. 

\begin{figure}[t]
\centering{
 \setlength{\tabcolsep}{5pt}
 \begin{tabular}{ccc}
  \includegraphics[height=0.4\linewidth]{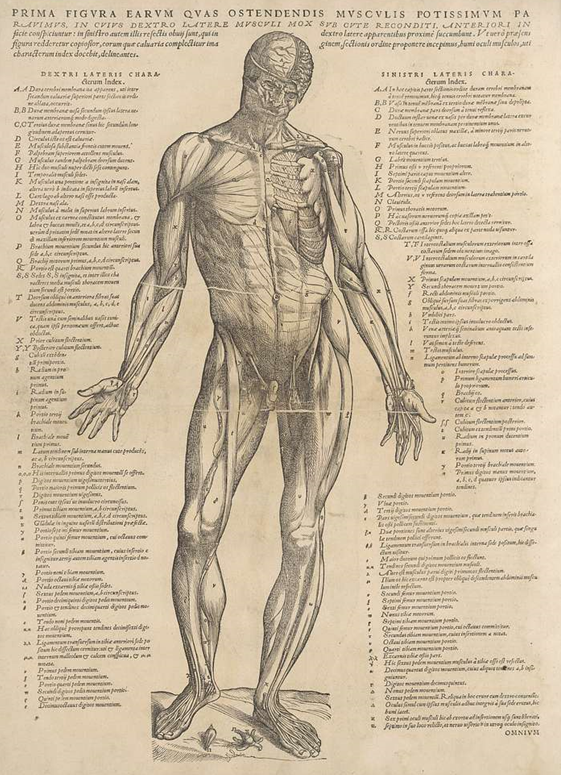} &
  \includegraphics[height=0.4\linewidth]{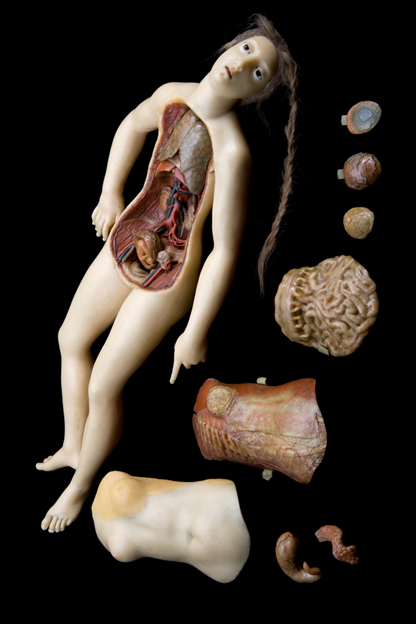} &
  \includegraphics[height=0.4\linewidth]{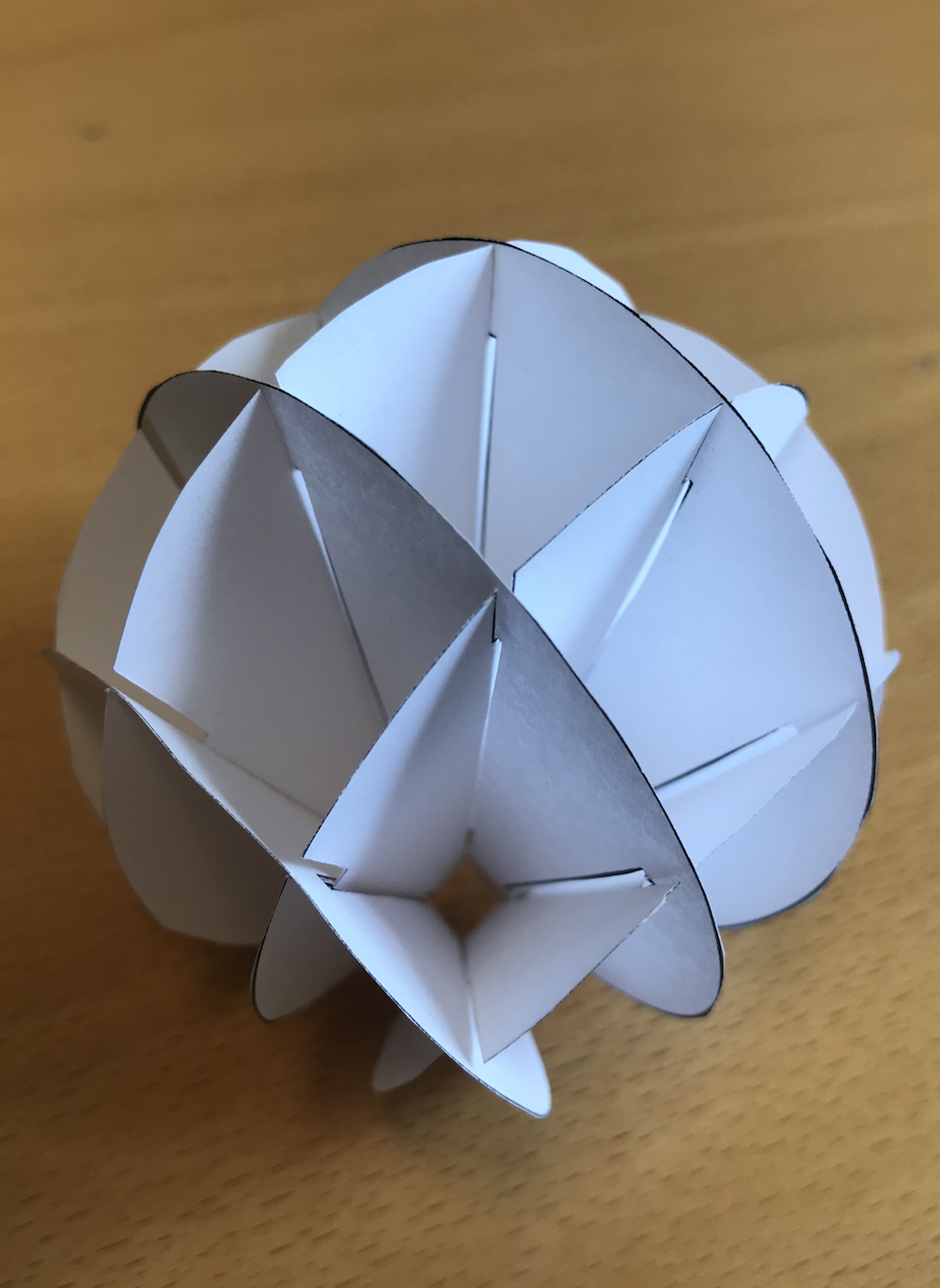} \\
   (a) & (b) & (c) \\
 \end{tabular}
}
\caption{(a) Illustration of the human muscular system, drawn in 1543 by A. Vesalius~\cite{andreas1humani}. (b)  An Anatomical Venus sculpture, produced at La Specola in c. 1785. (c) A sliceform of a sphere. \vspace{-10pt}
}
\label{fig:2_background}
\end{figure}

Hand-drawn anatomical illustrations appeared as early as 1522~\cite{choulant1852geschichte}. 
Figure~\ref{fig:2_background}(a) presents a well-known illustration of the human muscular system by A. Vesalius, drawn in 1543. 
People have been looking for alternatives to these 2D illustrations, for example by producing anatomical wax sculptures, such as the Anatomical Venus, shown in Figure~\ref{fig:2_background}(b). 
These 3D sculptures provided a reusable and durable anatomical education and training alternative to dissections. 
Later, the invention of medical imaging revolutionized the domain of anatomical education~\cite{preim2013visual}. 
Nowadays, as medical imaging techniques become increasingly sophisticated, many anatomical education applications have emerged~\cite{ blume2011google, smit2016online , halle2017open, preim2018survey}. 
These provide the immediate advantage of bringing on-screen personalized 2D or 3D representations of patients, in a time- and cost-efficient manner. 

Naturally, the use of physical representations of anatomical data has decreased. 
However, in recent years, the domain of data physicalization demonstrated that physical visualizations can be effective~\cite{moere2008beyond, Embodiment, jansen2015opportunities}---sometimes, even more than their virtual counterparts~\cite{jansen2013evaluating}. 
This is due to some of their unique properties, which are not encountered in traditional, screen-based visualizations, such as the sense of scale, or full engagement of the user’s perception and cognition. 
Medical applications within the domain of data physicalization revolve nowadays around the topic of 3D printing.
Still, 3D printing is expensive and cumbersome for private use~\cite{3Dprintingbasedonimagingdata}. 
At the same time, common 3D prints do not provide a lot of interactivity, due to their inherently static nature. 
Medical data physicalizations that do not involve 3D printing, but are made of affordable and accessible materials, are still limited~\cite{steinman2017data,Vol2velle, FraserBlocks}.

An additional consideration relates to the anatomical education of schoolchildren, where engagement is often sought by activity, such as playing, building, and crafting. 
Depending on the age, papercrafts, such as origamis, pop-ups, or sliceforms, are often used for educational purposes~\cite{eisenberg1998shop}.
An example of a paper-based sliceform of a sphere is shown in Figure~\ref{fig:2_background}(c).
Computer-assisted reconstruction of papercrafts has been investigated in the past, mainly for recreational purposes~\cite{le2013automatic, ruiz2014multi,paczkowski2018papercraft3d}. 
Crafting approaches have not been yet investigated for anatomical education, and they do not go beyond the use of other materials, except for paper (e.g., transparent). 

The \textit{contribution} of our work is the investigation of alternative forms of medical data physicalizations, which use readily available technologies and inexpensive materials.
We make use of the advantages of data sculptures (in the shape of semi-transparent sliceform crafts) within a cost-effective, simple, and easy-to-employ workflow. 
Our \textit{goal} is to support the creation of interactive and engaging anatomical physicalizations, to be used in laymen or children's education.
The main \textit{components} of our approach include: 
\begin{itemize} 
\item The support of both meshes and volumetric medical data.
\item An octree formulation and optimization for sliceforms to support volume data, based on a user-defined transfer function.  
\item A packing algorithm to relatively preserve an optimal assembly order and the volume structure while minimizing the used space.
\end{itemize}
\vspace{-2pt}

\section{Related Work}
\label{sec:related}

\noindent \textbf{Anatomical Education}---Preim and Saalfeld~\cite{preim2018survey} presented a survey on virtual anatomy education systems, focusing mainly on systems for medical students. 
A vast range of techniques is reviewed, spanning from surface and volume visualization~\cite{pommert2001creating, lichtenberg2016sline, vazquez2008interactive} to animations~\cite{bauer2014interactive}, and from anatomical labeling~\cite{ bruckner2005volumeshop} to virtual and augmented reality~\cite{saalfeld2016semi, pohlandt2019supporting, john2016use, Messier2016AnI3}.
The most influential examples include the VOXEL-MAN~\cite{pommert2001creating}, the open anatomy browser~\cite{ halle2017open}, and the Online Anatomical Human~\cite{ smit2016online}.
Applications for the general public include the ZygoteBody~\cite{ blume2011google} and the BioDigitalHuman~\cite{ qualter2012biodigital }. 
Preim and Saalfeld discuss \textit{constructivism}~\cite{huang2010investigating}, according to which active learning supports knowledge construction with less cognitive load, and \textit{embodiment}~\cite{jang2017direct}, according to which learning can benefit from the involvement of motorics and physical interaction.
This has been our starting point in the investigation of the suitability of data physicalization approaches for anatomical \textit{edutainment} (i.e., education and entertainment).

\noindent \textbf{(Medical) Data Physicalization}---Data physicalization can be summarized as the process of mapping data to objects and their properties of the physical world~\cite{moere2008beyond,jansen2015opportunities}.
The generated physical models aid or replace digital representations and allow data exploration with other senses, beyond the optical channel.
The physicalization of medical data has revolved, so far, mainly around 3D printing~\cite{ Hybrid3Dprinting, 3Dprintingbasedonimagingdata, CardiacBloodFlowPhys}.  
More complex models are even possible, due to the recent advancements of 3D printing, such as multicolor and polymaterial printing. 
A recent example includes the work of Ang et. al.~ \cite{CardiacBloodFlowPhys}, who developed a tangible physicalization of cardiac blood flow, to explore 4D MRI data in a slice-based physical model that complements traditional on-screen visualizations.
Despite its high educational value, 3D printing remains time-consuming, while it is not affordable. 
It also involves a complex workflow, which starts with the selection of an anatomical area, the creation of its 3D geometry, the optimization of the anatomical geometry for printing, and the selection of adequate printing technology and suitable materials. 
This makes it non-trivial for the general population, while the manipulation of material properties to mimic tissues is still not possible. 

Medical data physicalizations that do not involve 3D printing are limited. 
Except for the aforementioned wax models~\cite{markovicDevelopment}, other materials, such as wood, ivory, cardboard, and fabric, have been employed~\cite{olry2000wax}.
Notable are the papier-m\^{a}ch\'{e} anatomical models of Auzoux~\cite{papiermache}, which could also be taken apart for exploration.
Two more recent examples involve the creation of rearrangeable wooden models for the physical exploration of MRI brain scan cross-sections~\cite{FraserBlocks} and the usage of volvelles, which are interactive wheel charts of concentric, rotating disks~\cite{Vol2velle}. 
This physicalization mimics the on-screen fine-tuning of transfer functions to display meaningful volumetric data to non-knowledgeable users. 
Using paper, the production of volvelles is cost-effective and can be supported by home printers.

\noindent \textbf{Sliceforms and Papercrafts}---The topic of automatic creation of sliceforms and papercrafts has been tackled before, but not within the context of seeing through volumetric data or of an anatomical education application. 
For instance, there are automatic algorithms for generating stable, foldable, and physically plausible sliceforms, for designing and producing paper pop-ups from 3D meshes~\cite{ruiz2013generating, ruiz2014multi, xiao2018computational}.  Alternatively, origamic architecture papercrafts~\cite{le2013surface}, iris papercrafts~\cite{igarashi2016computational} or animated pop-ups that show motions of articulated characters~\cite{ruiz2015generating} are available.
Such approaches have been investigated for large-scale models~\cite{sass2016embodied, chen2017generative}, as well as for entire 3D scenes through developable surfaces~\cite{paczkowski2018papercraft3d}\XX{, and the automatic generation of paper architecture~\cite{li2010popup}}.
\XX{None of these approaches looks into generating physical models from volumetric data (only from mesh data), while they only employ paper.}

\vspace{-2pt}
 
\begin{figure*}
\centering{
\includegraphics[width=.95\linewidth]{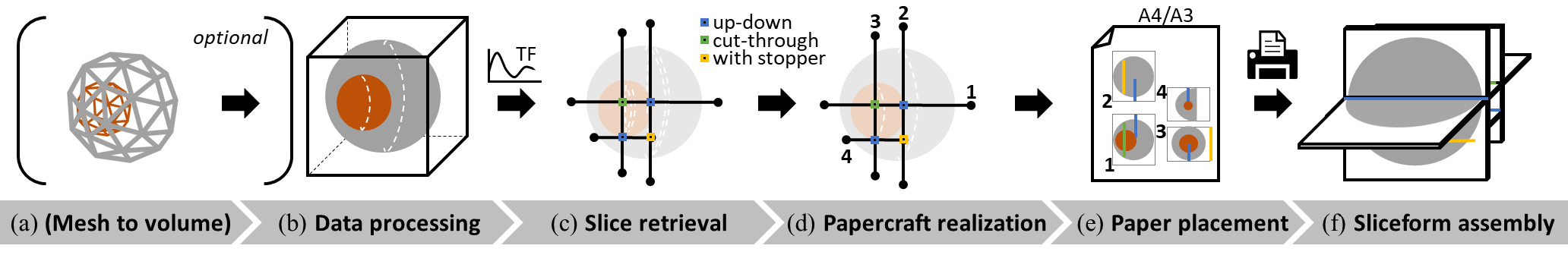}
}\vspace{-4pt}
\caption{The main steps of the \textit{Slice and Dice} workflow.
}
\vspace{-12pt}
\label{fig:3_method_pipeline}
\end{figure*}

\section{Requirements and Conceptual Choices}

The basic \XX{concept behind \textit{Slice and Dice}} is to provide a workflow for the cost-effective, simple, and easy generation of engaging physicalizations, which can be used in anatomical edutainment.
Our main motivation is that there is a lot of evidence that physical models can be effective in communicating data and processes~\cite{moere2008beyond, Embodiment, jansen2015opportunities}.
They have also been found to be more engaging than their virtual counterparts~\cite{jansen2013evaluating}.
At the same time, we want to take advantage of the so-called ``IKEA effect''~\cite{norton2012ikea}, according to which there is a tendency that people value things more if they have created them themselves.
To the best of our knowledge, such an approach does not exist \XX{in anatomical edutainment.} \noindent Our main \textit{requirements} were: \vspace{-4pt}
\begin{enumerate}
    \setlength{\itemindent}{10pt}
    \item [\textbf{(R1)}] The user has no (or little) knowledge of anatomy, and/or (physical) visualization.
    \item [\textbf{(R2)}] The workflow should accommodate anatomical meshes, and medical volumes.
    \item [\textbf{(R3)}] The user interaction with the workflow should be limited.
    \item [\textbf{(R4)}] The outcome of the workflow should be an easy-to-assemble and engaging physical model.
    \item [\textbf{(R5)}] The assembly of the physical model should be intuitive, time- and cost-efficient.
    \item [\textbf{(R6)}] The physical model should require easy-to-find and inexpensive materials, and common technology (e.g., home printers).
\end{enumerate}

\noindent Based on these requirements, our \textit{conceptual choices} were: \vspace{-4pt}
\begin{enumerate}
    \setlength{\itemindent}{10pt}
    \item [\textbf{(C1)}] \textbf{Supported data and functionality:} We consider the usage of both anatomical meshes and medical volumes. However, we do not focus on providing functionality for fine-tuning the visual properties of the employed surface or volume renderings, respectively. We limit ourselves to providing a number of anatomical mesh models, and a number of medical image volumes, with preset color and opacity (or transfer function) assignments.
    \item [\textbf{(C2)}] \textbf{Target physicalization:} We base our anatomical physicalizations on the generation of sliceforms. As we discussed in Section~\ref{sec:related}, sliceforms have been used before as a means of entertainment, but not as a means of edutainment. They are relatively popular and common among the general population, such as in greeting cards, 3D puzzles, and decorations. Sliceforms come also as a natural choice, given the usage of slice-based views for imaging data in medical practice.
    \item [\textbf{(C3)}] \textbf{Required materials and technologies:} For the creation of our anatomical physical models, we employ transparent films (acetate sheets), which are compatible with common printers, inexpensive, and easy-to-acquire at any stationery shop. Normal paper could also be used, but transparent films ensure visibility on the inner structures. An unintentional effect of the usage of transparent films is that the assembled sliceforms can have a pseudo-3D appearance.
\end{enumerate}

\vspace{-2pt}

\section{The \textit{Slice and Dice} Workflow}
\label{sec:method}

Figure~\ref{fig:3_method_pipeline} depicts the workflow of our \emph{Slice and Dice} strategy.
The workflow has been designed to primarily support volumetric data, but mesh data can also be employed.
Meshes are converted into a volume (Figure~\ref{fig:3_method_pipeline}(a)), before undergoing the remainder of the workflow.
The volume is initially processed and rendered on-screen, employing preset transfer functions (Figure~\ref{fig:3_method_pipeline}(b)).
Basic interactions, such as zooming, rotating, and panning, are possible.
Subsequently, the data is partitioned into slices, with the use of an octree that takes into account the visibility of the different structures, as conveyed by the preset transfer function of the volume rendering (Figure~\ref{fig:3_method_pipeline}(c)).
The slice configuration is then simplified, to minimize the number of separate planes that are used for the final sliceform assembly---thus, minimizing the assembly effort and time (Figure~\ref{fig:3_method_pipeline}(d)).
The slices are characterized based on their configuration within the octree structure.
These might be up-down slices (i.e., intersections), or cut-throughs (i.e., slices that are passing through one or more slices), or have a stopper (i.e., slices that ``hang'' from others).
Then, the assembly order is optimized to ensure that the sliceform can be assembled.
Finally, a packing algorithm is used to place the resulting slices, together with eventual annotations about the slice characterization and order onto the selected medium (Figure~\ref{fig:3_method_pipeline}(e)).
A home printer can be used, and after cutting, the slices are assembled into a sliceform to be explored (Figure~\ref{fig:3_method_pipeline}(f)).
All steps are discussed in detail, in the upcoming sections.
\vspace{-2pt}

\begin{figure*}
\centering{
 \setlength{\tabcolsep}{5pt}
 \begin{tabular}{c|c|c|c|c}
  \includegraphics[height=0.145\linewidth]{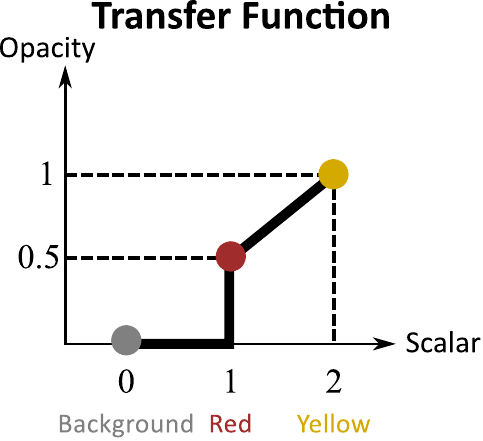} &
  \includegraphics[height=0.145\linewidth]{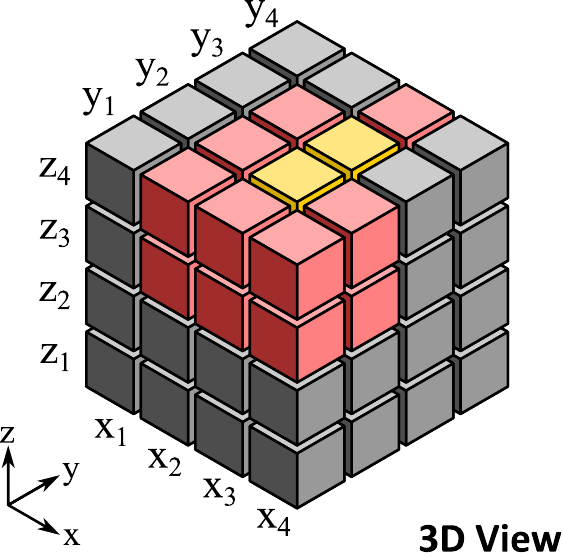} &
  \includegraphics[height=0.145\linewidth]{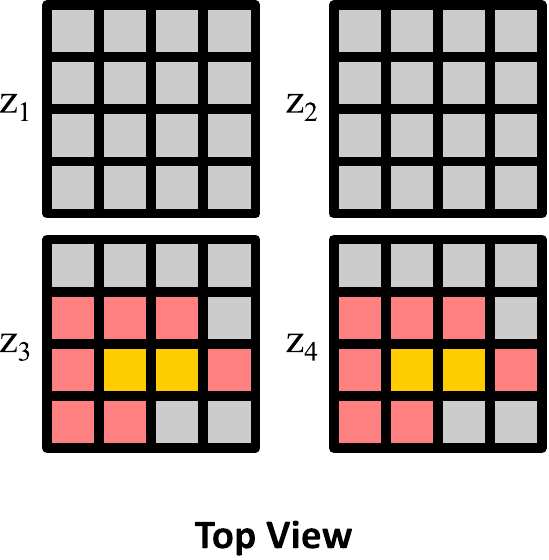} &
  \includegraphics[height=0.145\linewidth]{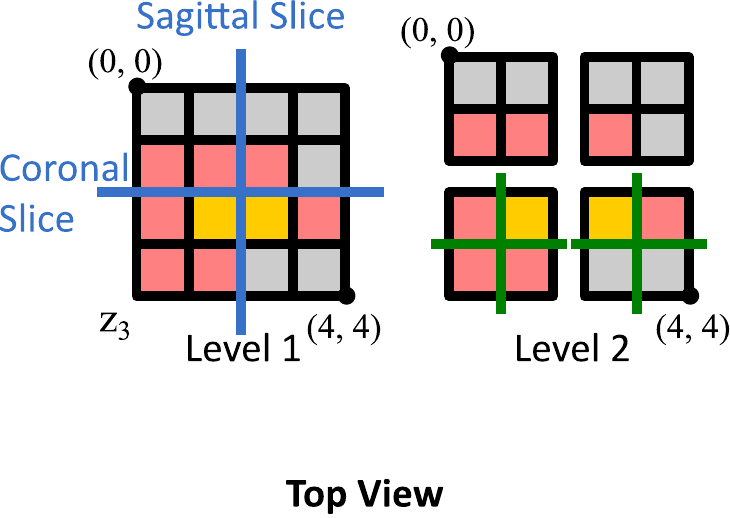} &
  \includegraphics[height=0.145\linewidth]{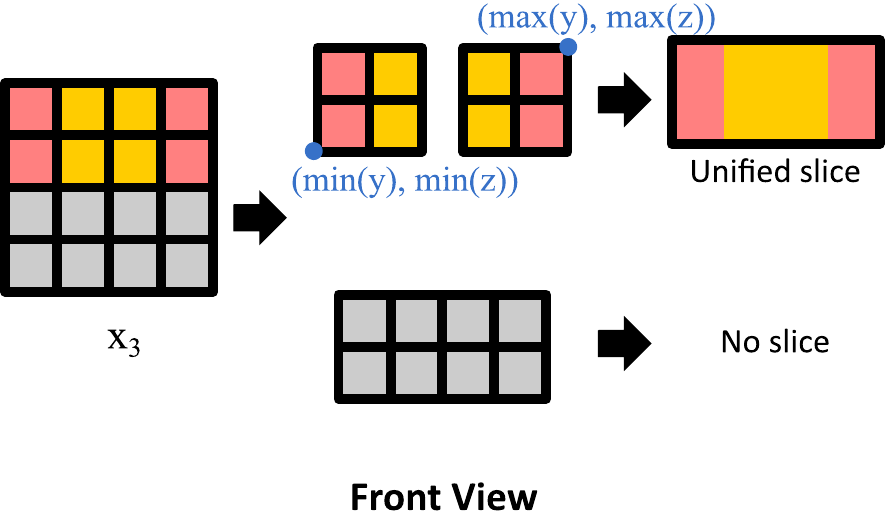} \\
   (a) & (b) & (c) & (d) & (e)\\
 \end{tabular}
}
\caption{An example of an octree-based spatial partitioning. (a) The transfer function of the volume. (b) The input volume. (c) Four axial slices (top view). (d) Two-level octree partitioning. (e) Unification of small adjacent slices.\vspace{-15pt}}
\label{fig:octree}
\end{figure*}

\subsection{Preprocessing and Rendering}
\label{sec:preprocessing}

The first step is to convert the data into a format that is usable by our workflow.
If mesh data are imported, these are converted into volumetric data, where each mesh is mapped to one intensity value.
Therefore, if a complex anatomical model of several meshes is loaded, each is treated as a distinct component, and is mapped to one distinct intensity value.
For the mesh data, we employ a traditional surface rendering for aesthetic reasons, and the converted volumetric data are used for computations only, from the next step on.
Instead, for the volumetric data, we employ a GPU-based volume rendering.
Basic interaction (zoom, pan, rotate) is supported.

The visual properties of this initial rendering are predetermined, and cannot be altered by the user.
There are three main reasons for this choice.
First, allowing the users to interact and change the transfer functions might be difficult, as they most probably have no knowledge of rendering, or of the underlying (medical) data.
Second, volumetric medical imaging data comes in many different formats and standards, and designing a transfer function that can accommodate all possible imaging options is impossible~\cite{ljung2016state}.
Third, there is a vast literature on transfer function design~\cite{ljung2016state}, and strategies for visibility in rendering~\cite{viola2005smart, bruckner2006illustrative, hadwiger2003high}, which cannot be all included.
Hence, we consider the tuning of transfer functions out of scope for this work, and we provide a preset of volumetric data with predetermined transfer functions.
For the mesh data, we assign opacities based on the location, where inner structures have higher opacity, and colors are assigned in a semi-random way so that each structure is discernible from the rest.
\XX{However, in the future, it would be interesting to investigate optimization strategies for the employed transfer functions in the physicalizations, as well as the effect of their superimposition (and order) within the physical model.}

\subsection{Slice Retrieval}
\label{sec:octree}
The next step in our workflow is to generate representative slices from the available volumetric data.
Lattices are used to describe strips of materials crossed and fastened together.
Mathematically, lattices have interesting properties that can be used to investigate and formalize the structure of objects, in two or three dimensions.
For this reason, lattices are used in sliceform papercrafts, to better represent the features of a target object.
In contrast to classical sliceform designs, which try to maximally preserve the features of the outer surfaces, in this paper, we aim to find a lattice partition that better preserves the inner structures of our target volume.
In computer graphics, an octree is a space partitioning technique for fast searching neighbors, using a tree data structure.
By recursively subdividing the input 3D space into eight octants, an octree allows to hierarchically manage sparse and coarse information, in the 3D space.
In this paper, we are inspired by the concept of an octree, and introduce an \textit{octree formulation for sliceform papercrafts of volumetric data}.
After the octree-based spatial partitioning, we optimize the slice configuration by \textit{unifying adjacent slices} to form bigger ones.
This is done to facilitate, and speed up the assembly.

\noindent \textbf{Octree-Based Volume Partitioning}---Classical octrees are built for searching in a fast way which objects are nearby in a 3D space; for example, points of a mesh that are nearby in a 3D scene~\cite{Brunet:1992:octree}.
The algorithm begins with a root node, and subdivides iteratively each tree node into eight children (octants)---if and only if, dense information is stored within the node region.
In our approach, we replace this conditional density statement by the total number of quantized voxel types within the corresponding tree node.
This replacement is done, because each quantized value in the transfer function reflects a distinct structure in a medical dataset, which we consider an interesting sub-region that requires more slices.
The ``importance'' of the quantized voxels can be retrieved from the transfer function (Figure~\ref{fig:octree}(a)).
To determine which structures are more ``important'', the scalar value of a voxel within the volume data is multiplied by its visibility factor (e.g., its opacity), as given by the transfer function.
For example, if the transfer function of a CT volume dictates that all intensity values of $1000$ have $0.5$ opacity, the scalar value $1000$ will be scaled by $0.5$.
Therefore, values that are more opaque are ``prioritized'' in our approach, and transparent values (i.e., hidden structures) are not considered.
Figure~\ref{fig:octree}(b,c) shows an octree partitioning example, where three quantized values are included in the volume.
Gray color is considered as the background (opacity $= 0$) and is ignored in our calculation.
Red and yellow voxels indicate different structures in the dataset (opacity $=0.5$ and $1$, respectively).
In Figure~\ref{fig:octree}(d), our octree algorithm subdivides the entire volume (Level 1), and recursively subdivides the sub-volumes until no different features exists, or at the finest voxel resolution (Level $2$, in this example).
We also introduce a parameter $L$ that allows users to specify the highest tree level, in order to control the complexity of the final papercraft model.
However, with an octree, we may generate several small slices adjacent to each other, which complicate the assembly process.
Next, we explain how we integrate those slices into bigger, for easier assembly.

\noindent \textbf{Slice Generation and Unification}---In our papercraft design, we take two perpendicular slices, starting from the center of each tree node; for instance, one coronal ($yz$ plane) and one sagittal ($xz$ plane) slice (Figure~\ref{fig:octree}(d)).
This gives our papercraft flexibility to fold the 3D model into 2D, for better portability.
We exclude the bounding planes of each tree node, to avoid having slices on the same 2D plane---especially, if the tree nodes belong to different tree levels.
With this choice, when two slices share a boundary edge, they might be still distinct from each other.
Figure~\ref{fig:octree}(e) shows such an example.
If we consider the coronal slice $x_3$ in Figure~\ref{fig:octree}, the upper half is subdivided twice, while the lower half is subdivided once.
With an octree, we will receive two separate small slices: two subdivided half-red and half-yellow slices.
For these adjacent slices, the assembly will be tedious.
We, therefore, combine these slices into one, by computing their minimum and maximum coordinates to synthesize their unification.
Note that this is done by first searching the connected slices on the same 2D plane, and then performing the unification process, to find the appropriate number of rectangular slices that covers the entire extent on the same 2D plane.
\vspace{-2pt}


\subsection{Papercraft Realization}
\label{sec:order_optimization}

The quality of a papercraft does not only rely on the feature representation of the input data.
It also depends on the feasibility of its realization, based on slice intersections and order of assembly.
To support the realization of the sliceform assembly, we first \textit{categorize all pairwise slice intersections}, as hinge types within the papercraft. Then, we \textit{determine a reasonable hinge order}, to guide users through easier assembly.

\begin{figure}[b]
\centering{
 \setlength{\tabcolsep}{5pt}
 \begin{tabular}{ccc}
  \includegraphics[width=0.25\linewidth]{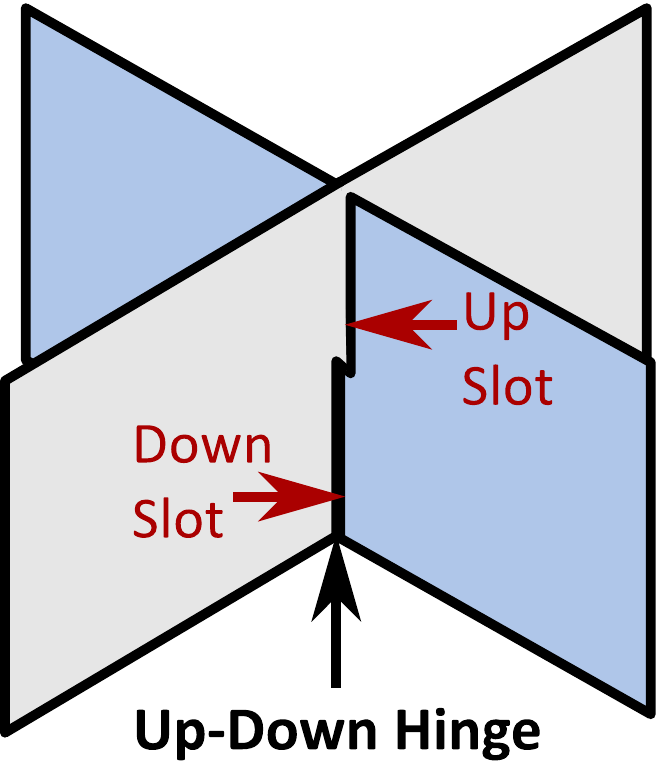} &
  \includegraphics[width=0.25\linewidth]{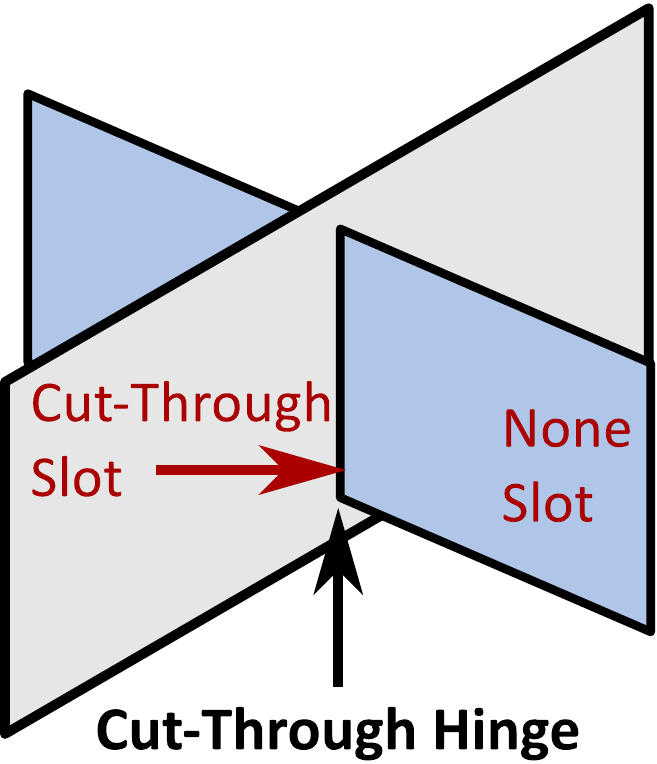} &
  \includegraphics[width=0.25\linewidth]{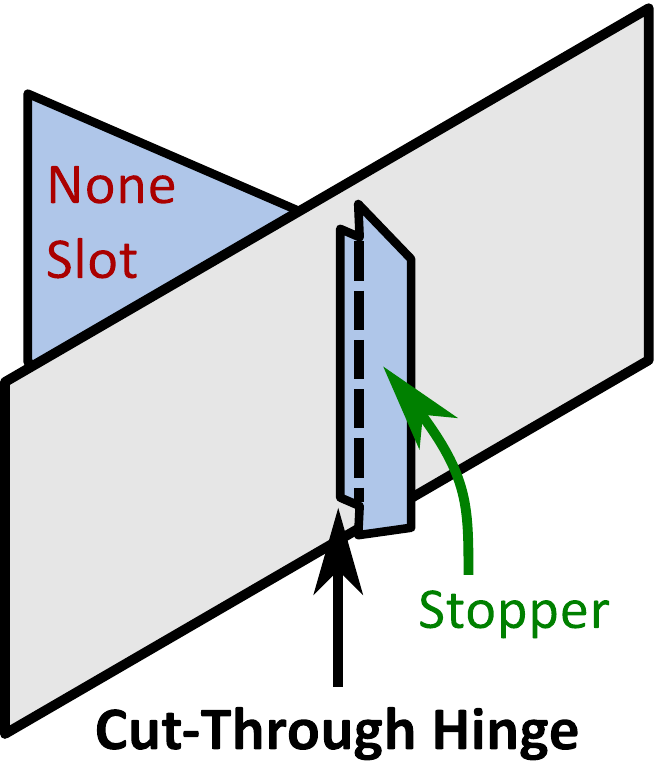} \\
   (a) & (b) & (c)\\
 \end{tabular}
}
\caption{Types of hinges: (a) an \emph{up-down} (intersection) hinge, and \emph{cut-through} hinges (b) without, and (c) with a stopper.}
\label{fig:hinge}
\end{figure}

\noindent \textbf{Hinge Characterization and Construction}---To realize the sliceform papercraft, we first need to find a way to stitch (i.e., intersect) two slices together.
As introduced by Le-Nguyen~et~al.~\cite{le2013automatic}, each line segment in the intersection of two slices is called a \textit{hinge}.
Two types of hinges are commonly used in a sliceform model: \emph{up-down intersections} and \emph{cut-through} hinges.
For \emph{up-down} hinges, we can cut the upper hinge segment of the coronal slices and the lower hinge segment of the sagittal slices, and stitch the two slices by moving them vertically along the hinges (Figure~\ref{fig:hinge}(a)).
The cut-away upper or lower hinge segment is called a \textit{slot}.
For \emph{cut-through} hinges, the cut-away hinge segment is located on the bigger slice along the $z$-axis, and no cutting will be performed on the smaller one (Figure~\ref{fig:hinge}(b)).
Its corresponding slot is called a \emph{none} slot, since no cut will be performed, and it will be inserted through the \emph{cut-through} slot.

\noindent \textbf{Assembly Order Optimization}---\XX{Our assembly order concept stems from common practices in 3D anatomy puzzle games, where small, inner structures need to be assembled, before large, outer ones.
However,} the feasibility of a sliceform depends on the condition that there are no interlocks in the assembly order, to facilitate the reconstruction.
\XX{The idea of an outwards assembly is less rigorous (up to a certain degree) since it primarily influences user experience in assembly. The feasibility of reconstruction, though, needs to be formulated by hard constraints; otherwise, the physical model cannot be assembled.
These observations can be summarized as:}

\begin{enumerate}
\setlength{\itemindent}{10pt}
\item[\textbf{(O1)}] \textbf{Order Inequality:} The order of each hinge is distinct, since we assume that hinges are stitched sequentially.
\item[\textbf{(O2)}] \textbf{\emph{Cut-Through} Hinge Dependency:} Given a slice with a \emph{none} slot between two \emph{up-down} hinges, the \emph{cut-through} hinge should be assembled earlier than the \emph{up-down} ones.
\item[\textbf{(O3)}] \textbf{Backbone Hinge:} The coronal and sagittal slices of the root node in the octree form the backbone hinge. This is the longest hinge with the largest slices. The backbone hinge is expected to be stitched first, to maintain the papercraft stability.
\item[\textbf{(O4)}] \textbf{Radial Dependency:} The assembly of hinges on a slice should be composed starting from the center.
\end{enumerate}

\noindent These statements led us to consider the assembly order problem as a constrained permutation problem, where each element in the set should be distinct, and certain partial orders in the set should be preserved.
We formulate this problem as an integer programming problem, where \textbf{(O1)}--\textbf{(O3)} are hard constraints and \textbf{(O4)} is a soft constraint.
The reason for choosing integer programming rather than another optimization technique is that it allows us to separate hard and soft constraints.
It is also capable of finding the optimal solution, instead of a feasible solution, if the scale of the problem is relatively small.
Our input in the formulation are the indices of the hinges $H=\{1,2,...,n\}$, and the output will be a sequential order of hinge indices.
Mathematically, the \textbf{(O1) Order Inequality} is equal to $x_i<x_j$ or $x_i>x_j$, and can be formulated as:
\begin{eqnarray}
\label{eq:inequality}
  \begin{array}{llcl}
  \forall i,j \in H~\textrm{and}~i \neq j, & x_i +1 &\leq& x_j + M \cdot \alpha_{ij}       \\
                     & x_j +1 &\leq& x_i + M \cdot (1-\alpha_{ij}),   \\
  \end{array}
\end{eqnarray}
where $x_i$ and $x_j$ correspond to integer hinge indices. $\alpha_{ij}$ is a binary variable for validating one of the above equations in the solution space. It is applied together with a large constant $M$. $M$ in our setting should be at least larger than the total number of hinges.
The \textbf{(O2) \emph{Cut-Through} Hinge Dependency} is formulated as:
\begin{eqnarray}
\label{eq:ct-dependency}
  \begin{array}{llcl}
  \forall i,j,k \in C~\textrm{and}~i \neq j \neq k,
                     & x_j +1 &\leq& x_i         \\
                     & x_j +1 &\leq& x_k,        \\
  \end{array}
\end{eqnarray}
where $C$ collects all triples $(i, j, k)$ where $i, k$ are \emph{up-down} hinges, and $j$ is a \emph{cut-through} hinge, and hinge $j$ is adjacent to hinges $i$ and $k$ on a slice.
The incorporation of \textbf{(O3) Backbone Hinges} is intuitive and can be formulated as $x_0 = 0$, since we force to place the backbone hinge $x_0$, in the first place.
As for the soft constraint \textbf{(O4) Radial Dependency}, we formulate it as the objective function in our optimization:
\begin{eqnarray}
\label{eq:cost}
 objective_{O4} = \sum_{x_i \in H}{w_{distance}(i) \cdot x_i},
\end{eqnarray}
where $w_{distance}(i)$ denotes the Euclidean distance of a hinge to the center of the volume.
Once we retrieve the order of hinges, we use this order to guide the users, to assemble the papercraft.
Two instructions can be provided.
One is the order of hinges, and the other is the order of slices, which follows the order of hinges.
\vspace{-2pt}

\subsection{Paper Placement}
\label{sec:paper_placement}

Now that we have the optimal ordering for the assembly of the sliceform, we need to put the distinct slices onto a piece of transparent film of a user-selected format (A3 or A4).
For this, we need to take into account three considerations:
First, slices that are ``nearby'' in the 3D volume space (and, subsequently, in the octree) need to be ``nearby'' on the printable configuration \textbf{(P1)}.
This is to support the intuitive assembly of the sliceform.
Second, the optimized order of assembly needs to be respected \textbf{(P2)}.
This is to support the easy assembly of the sliceform, given that there might also be cut-through slices (with or without a stopper), which are more complex to put together.
Third, we have to make the best possible use of the given materials (i.e., optimal use of the size of the transparent film) \textbf{(P3)}.
This is to guarantee that the object is big enough to be explored, but also fits within the available sizes.

To this end, we construct for each slice a 4D vector $\vec{v} = (x, y, z, d)$ that contains the normalized coordinates in the 3D space of the center $(x,y,z)$ of the slice \textbf{(P1)} and the order $d$ \textbf{(P2)} of the assembly.
Normalization ensures that all variables are weighted the same within the vector, which is important if our approach is based on Euclidean distances.
For example, if we have five slices ($d \in [0,4]$) within an $\mathbb{R}^3$ space $[-100,100]^3$,  without normalization, the coordinates $(x,y,z)$ and the order $d$ will be given unequal weights depending on their variance.
Principal Component Analysis (PCA)~\cite{pearson1901liii} is done in order to find if any variables of the vector $\vec{v}$ might be correlated, to get rid of redundancy and to identify the two components that maximize variance.
Then, a $k$-means clustering~\cite{lloyd1982least} partitions the slices into $k$ clusters.
Each slice belongs to the cluster with the nearest mean.
For calculating the optimal number of clusters, we employ the elbow method~\cite{thorndike1953belongs}.
The clustering gives an indication of slices that are located ``nearby'' in the 3D space \textbf{(P1)}, and in order of assembly \textbf{(P2)}.

The users can select the paper format, and whether to use one or more sheets.
To make optimal use of the available printing material \textbf{(P3)}, we apply the following packing approach.
First, a $k$d-tree~\cite{bentley1975multidimensional} is applied to partition the available paper space based on the results of the previous clustering, while taking into account the sizes of each of the slices.
If a cluster contains more (or bigger) slices, its respective paper partition will be bigger to fit all slices.
The sizes of the partitions are constrained so that all partitions fit within the available paper format.
This step results in a number of partitions equal to the number of clusters and each of these partitions contains the slices of the respective cluster, which are uniformly scaled to preserve their proportions.
The next step involves taking a rectangle bin packing algorithm~\cite{jylanki2010thousand} to determine how the slices should be packed into the partitions, as well as their optimal sizes, to cover as much as possible space in the available paper format.
Rotations of slices within the bins are possible.
We use a Maximal Rectangles algorithm with the best short side fit~\cite{jylanki2010thousand}.
The algorithm stores the slices into the partition that they belong in such a way that the length of the shortest leftover side of the partition is minimized.
Thus, $min(w_f-w, h_f-h)$ is the smallest, where $w_f$ and $h_f$ are the partition dimensions, and $w$ and $h$ are the slices dimensions.
Once the packing is completed, the partitions are assembled into the selected size of the paper and the number of sheets.
\vspace{-5pt}

\subsection{Papercraft Stability}
\label{ssec:stability}
There are several factors that influence the stability of the final papercraft.
Some essential factors include \textbf{(S1)} the weight balance of the octree partition, \textbf{(S2)} unexpected easy-to-open hinges, and \textbf{(S3)} the materials used for the papercraft.
In rare cases, if the features of a volume are only located at one of the top corners, which requires a large number of slices for appropriate representation, the weight balance of the papercraft can be skewed.
This did not happen in our experiments, since we normalized the volume coordinates, after reading the data.
Nevertheless, one can sum up the gravitational torques of slices to test if the current model is weight-balanced, or add additional ``empty'' slices at the bottom, to compensate for this issue.
Since our algorithm is based on octree partitioning, we may encounter a case, where a \emph{none} slot is located exactly at the boundary of the slice. For this, we introduce a stopper, as shown in Figure~\ref{fig:hinge}(c).
In our experiments, we tested our papercraft using copy papers, cardstock papers, and acetate sheets (transparent films). All paper types worked, but the slots should have a minimum width (e.g., $1$ mm) to facilitate the sliceform assembly.

\subsection{Print and Assembly}
For the printing, the user can choose between A3, A4, and a user-determined size, and the partitions can be printed on one or multiple papers, depending on how big the physical model should be.
The assembly printout contains labels and annotations for the slices, denoting the order in which the slices should be assembled, slots for the up-down and cut-through hinges, as well as stoppers.
We include also a rendered 3D model of the data, annotated with the octree cuts, and the order of slices, to facilitate the papercraft assembly.
For the cutting, a cutting plotter can be used, but simple scissors, or a utility knife, are sufficient, and more accessible.





\section{Results with Mesh Datasets}
\label{sec:mesh_results}

\begin{figure*}[!t]
\begin{minipage}{0.67\textwidth}
\centering{
  \setlength{\tabcolsep}{3pt}
 \begin{tabular}{ccc}
  \includegraphics[height=1.15in]{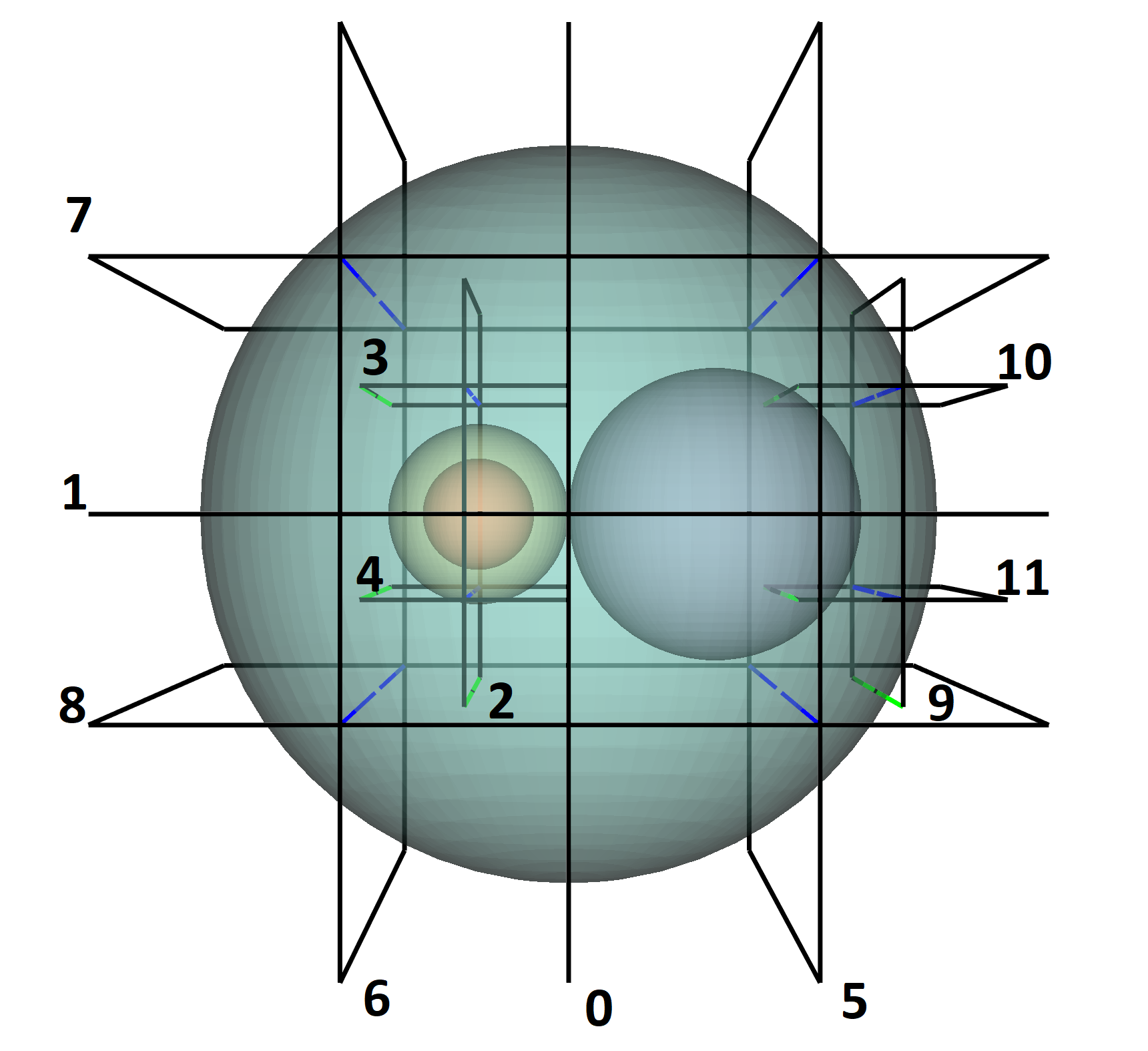} &
  \includegraphics[height=1.15in]{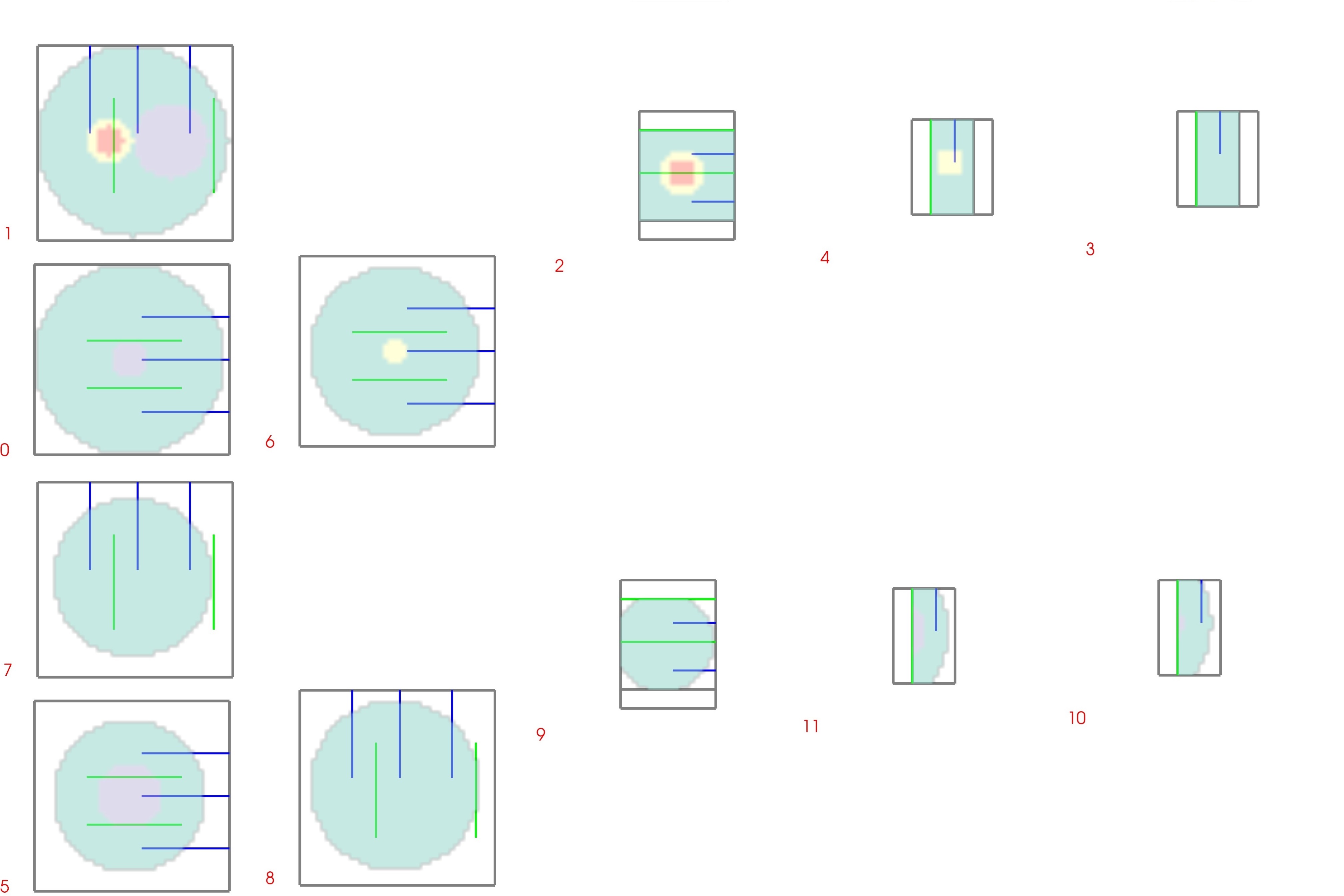} &
  \includegraphics[height=1.15in]{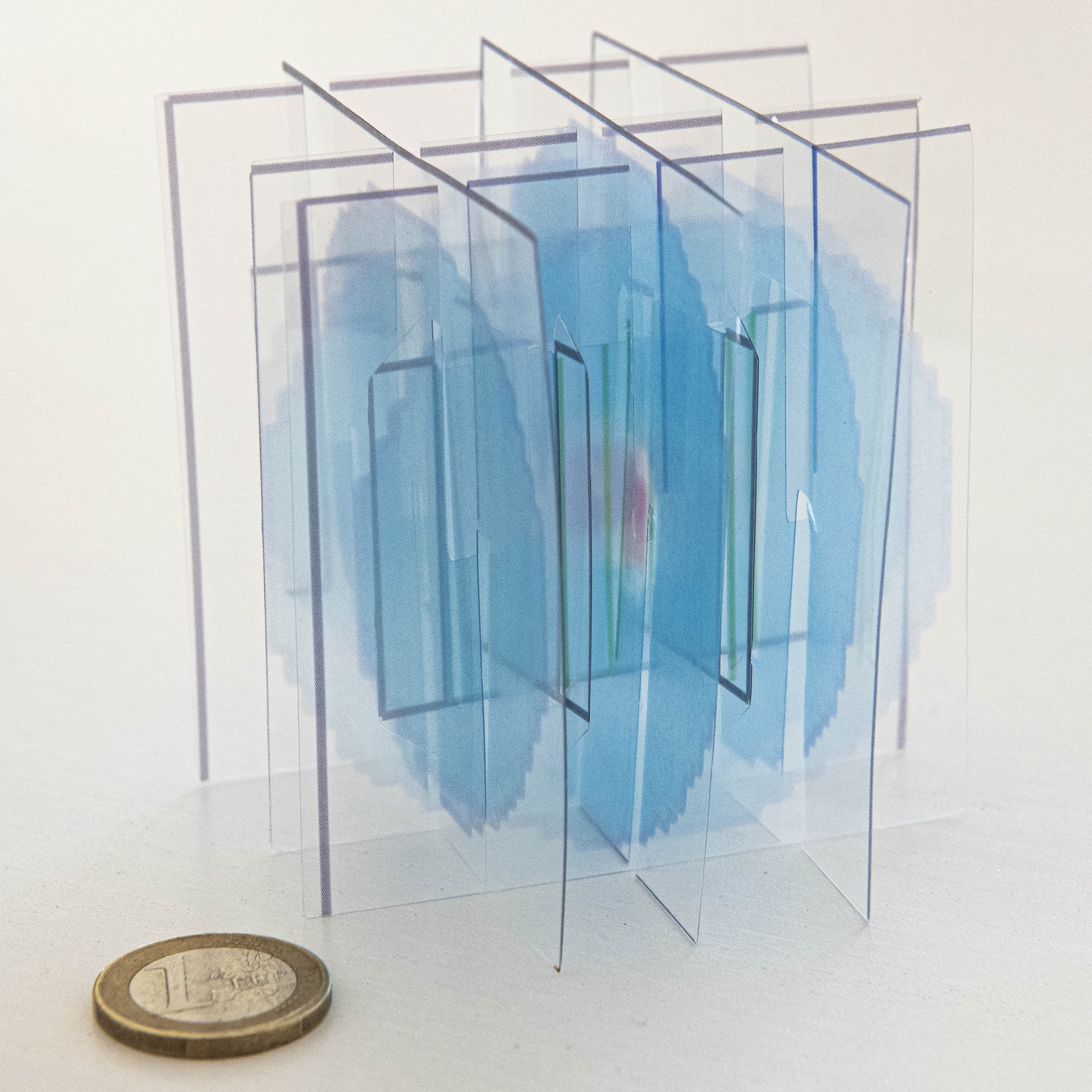} \\
   (a) & (b) & (c)\\
  \includegraphics[height=1.15in]{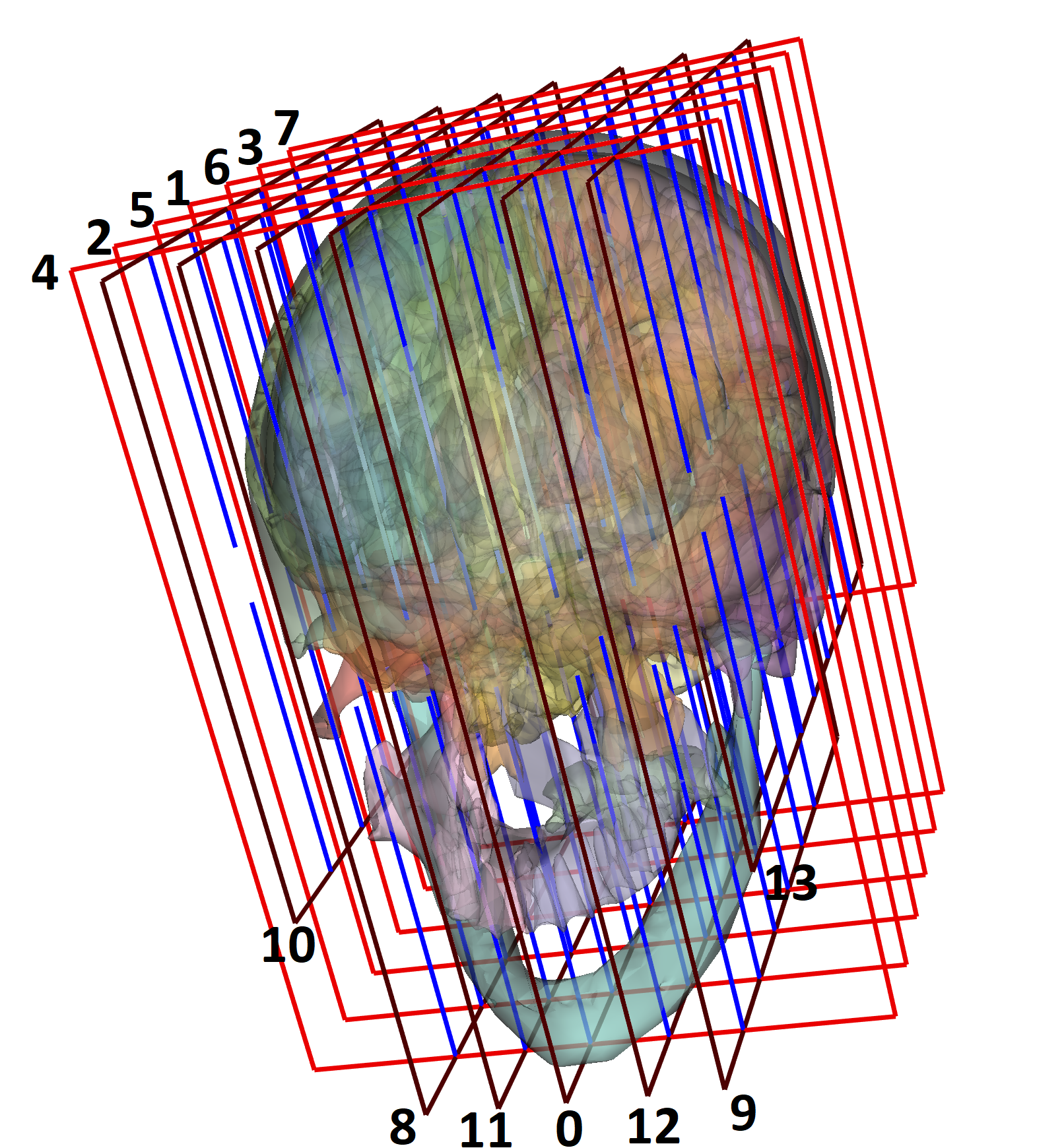} &
  \includegraphics[height=1.15in]{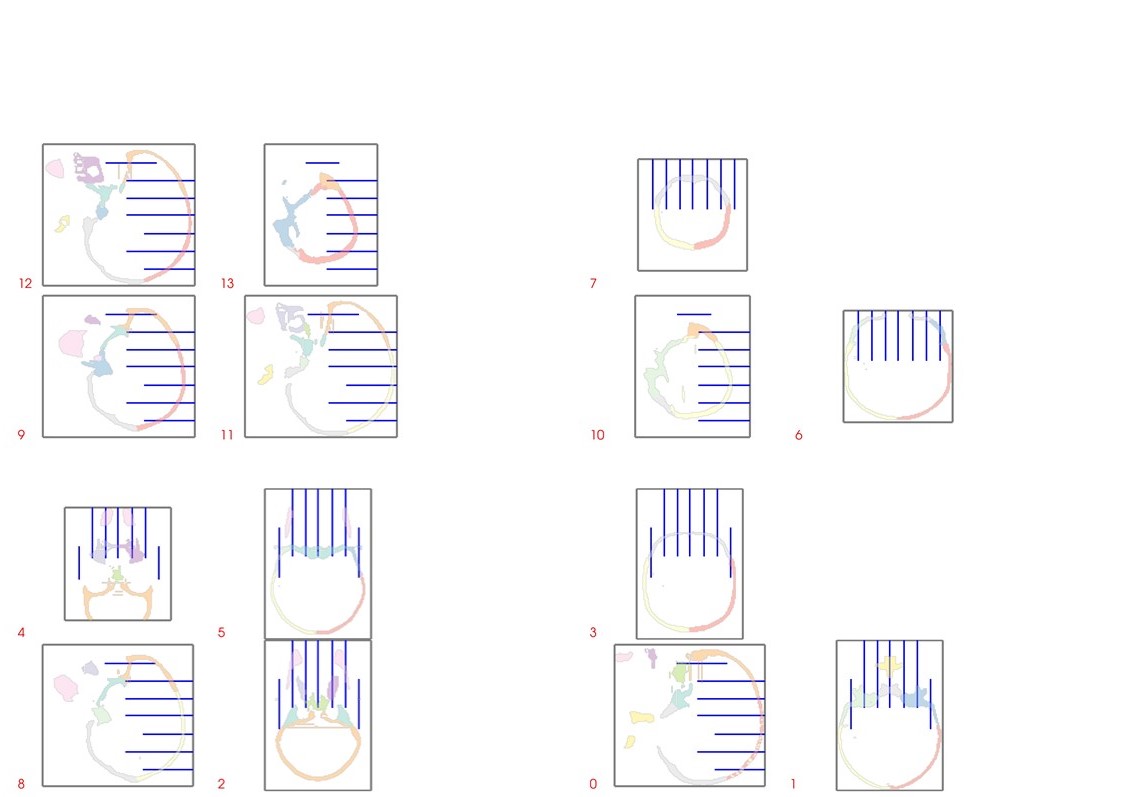} &
  \includegraphics[height=1.15in]{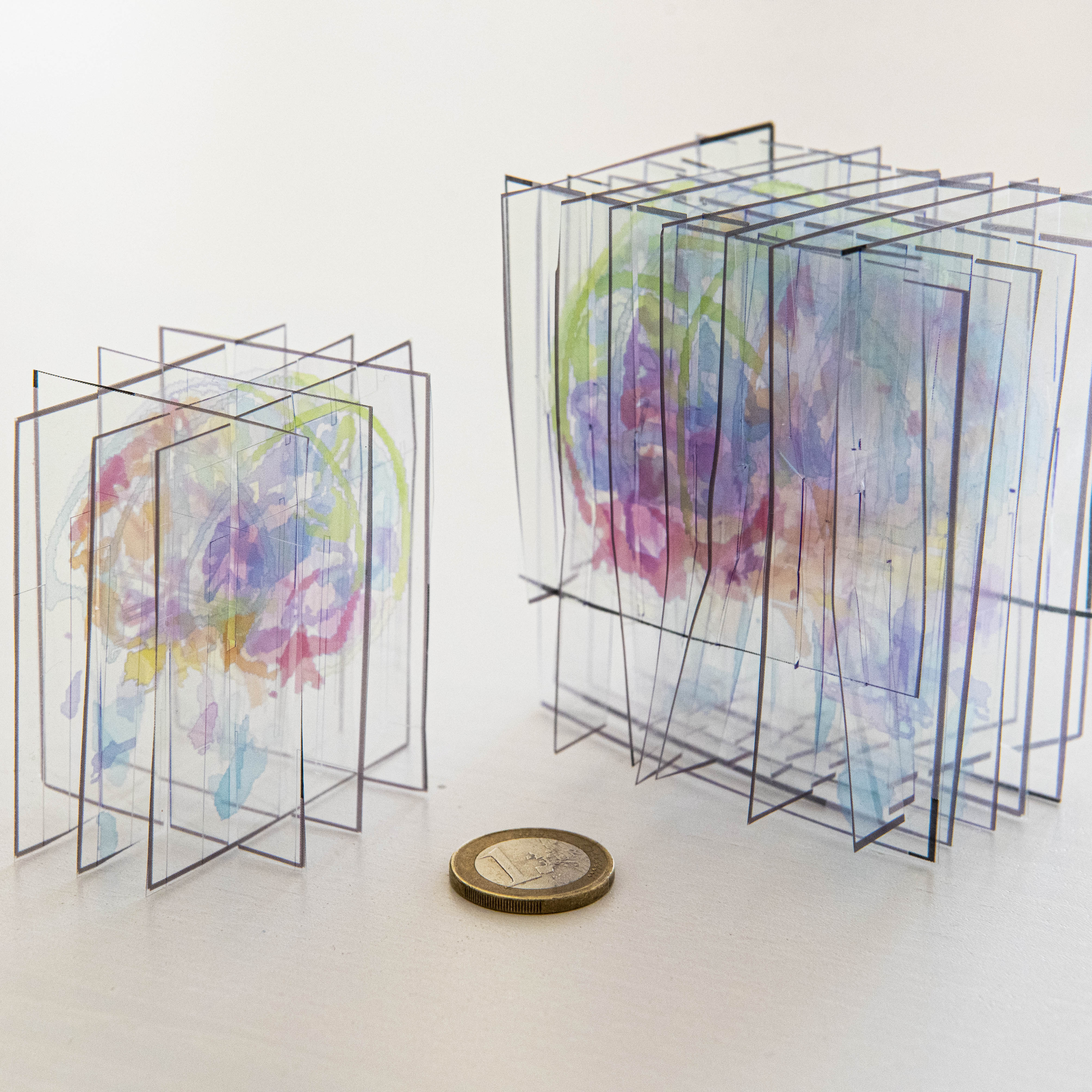} \\
   (d) & (e) & (f)\\
  \includegraphics[height=1.15in]{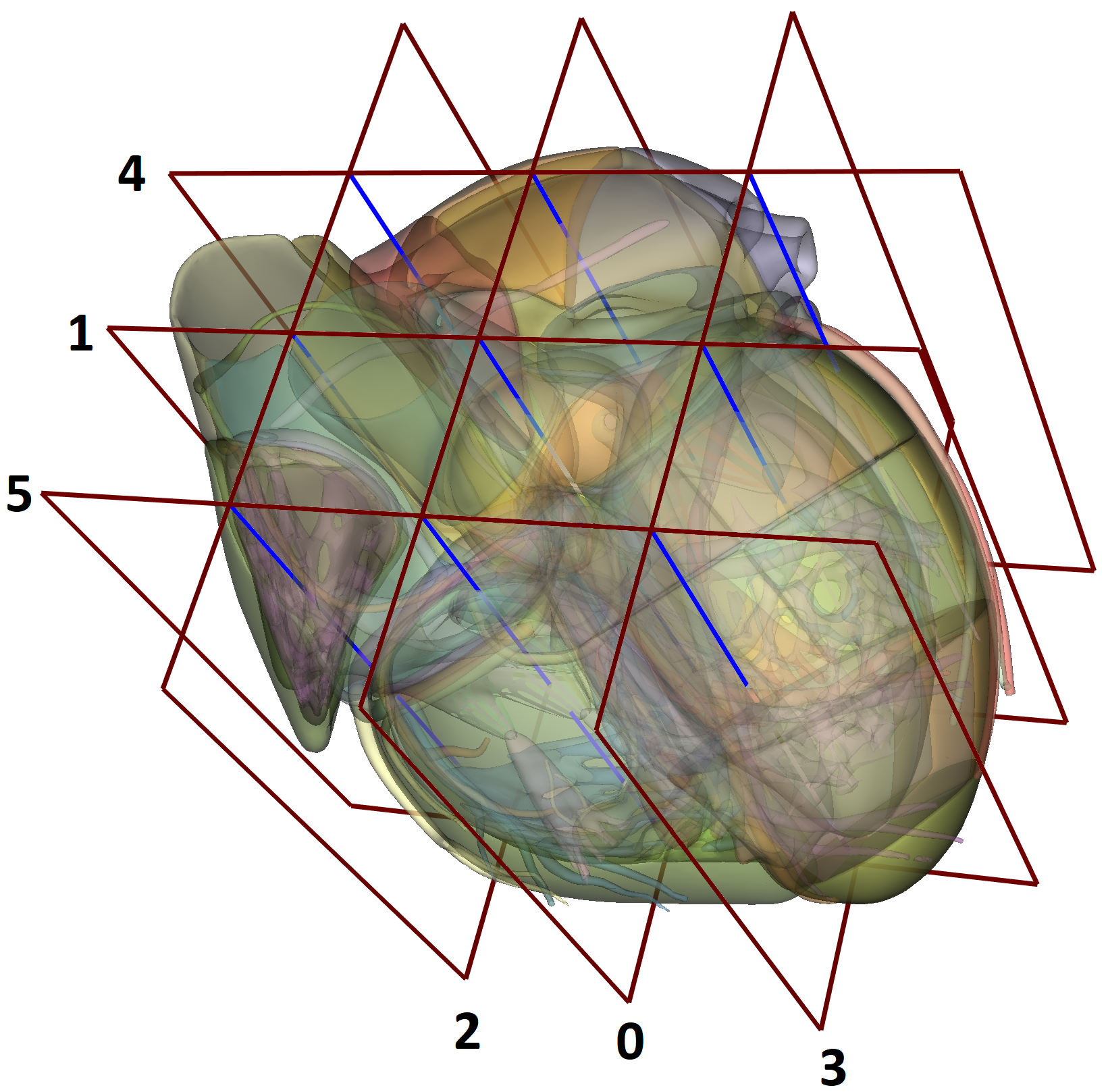} &
  \includegraphics[height=1.15in]{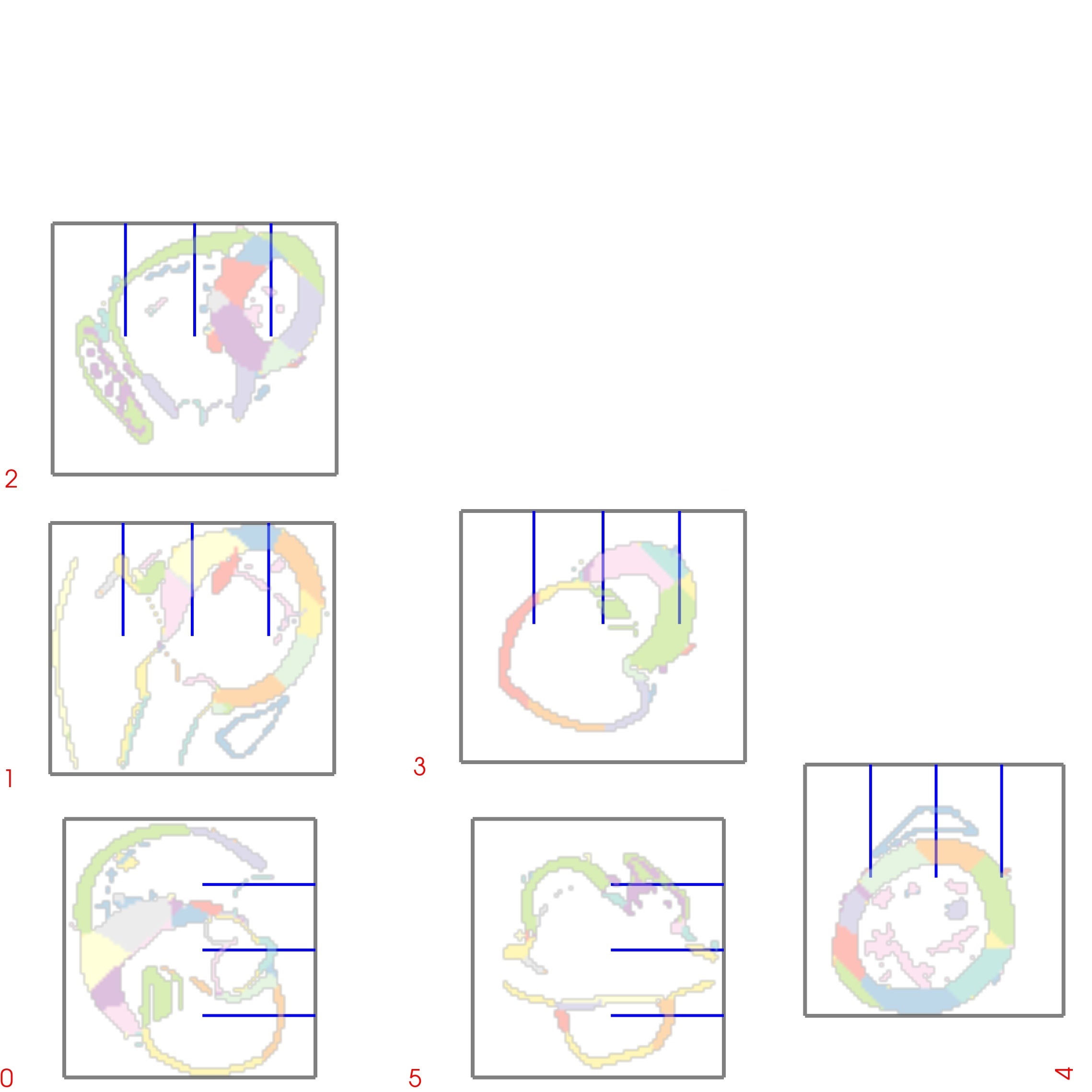} &
  \includegraphics[height=1.15in]{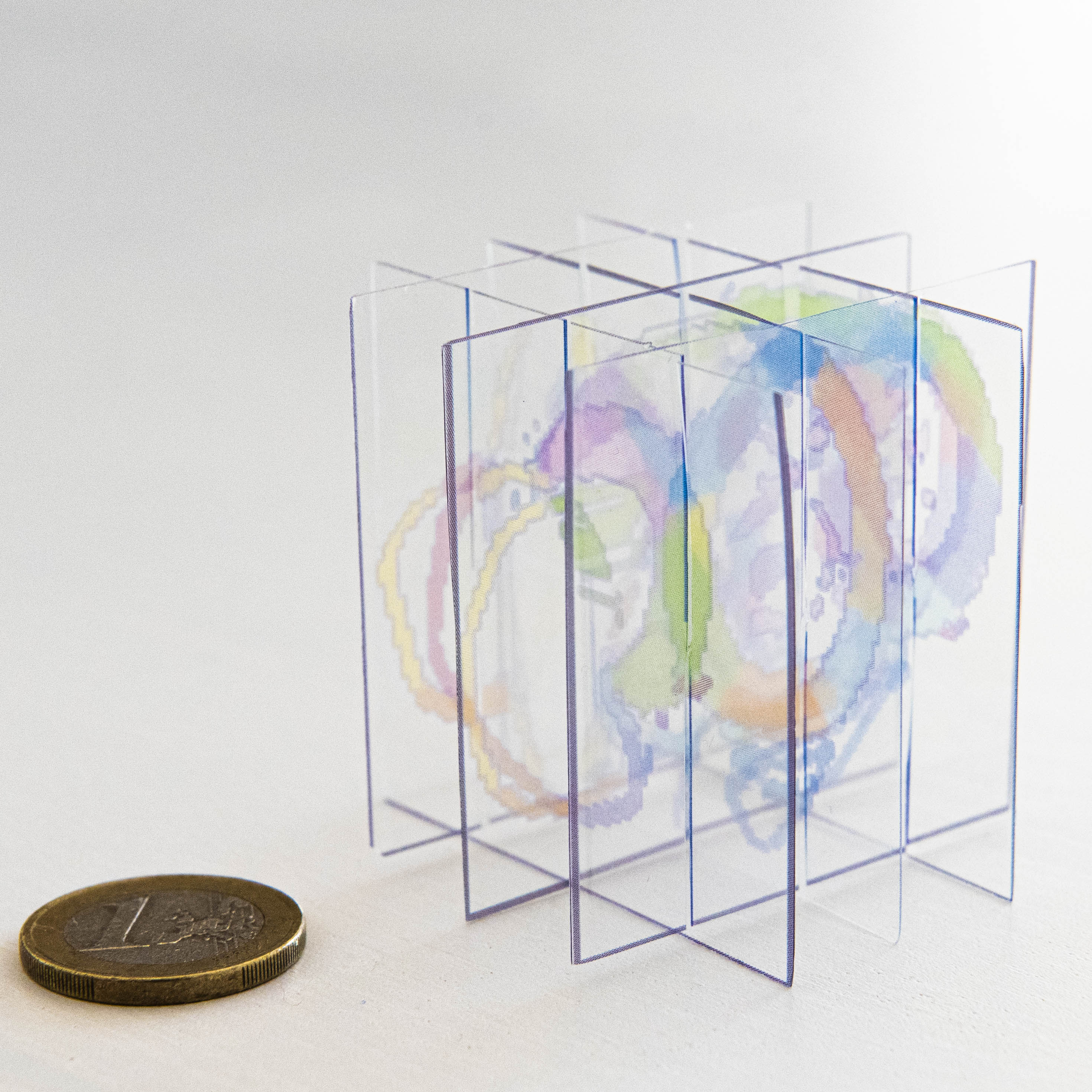} \\
   (g) & (h) & (i)\\
 \end{tabular}
}
\caption{Results for three mesh datasets. In (a)--(c), we show a synthetic dataset, with a configuration of four nested spheres. In (d)--(f), we present a head mesh dataset, and in (g)--(i) a heart mesh dataset. The head and heart meshes are from the \textit{BodyParts3D} database~\cite{mitsuhashi2009bodyparts3d}. \vspace{-10pt}}
\label{fig:result_mesh}
\end{minipage} %
\begin{minipage}{0.01\textwidth}
\parbox{0.01\textwidth}{}
\end{minipage} 
\begin{minipage}{0.3\textwidth}
	\centering{
	  \setlength{\tabcolsep}{1pt}
 \begin{tabular}{c}
  \includegraphics[height=1.2in]{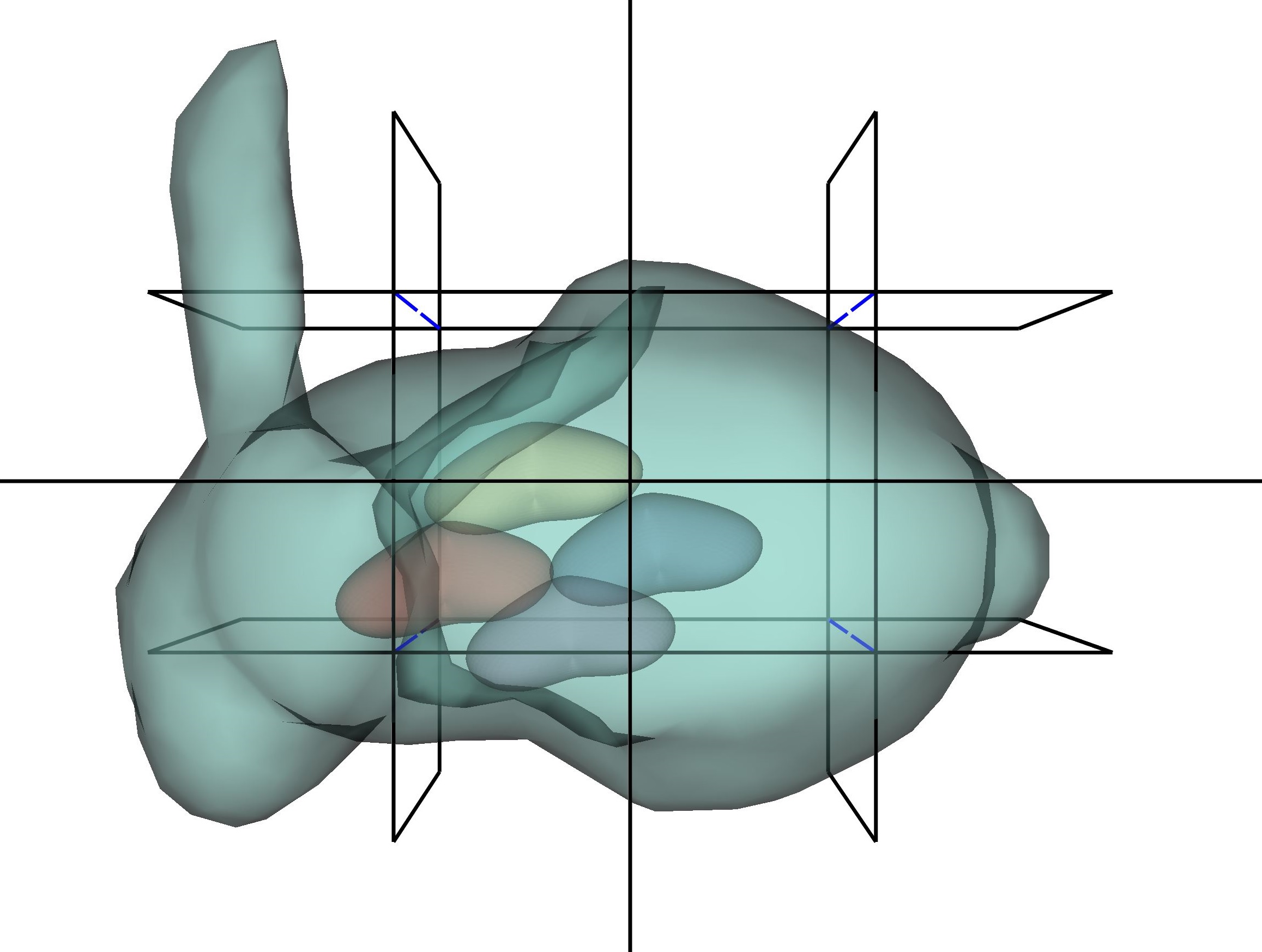} \\
   (a) $L=2$\\
  \includegraphics[height=1.2in]{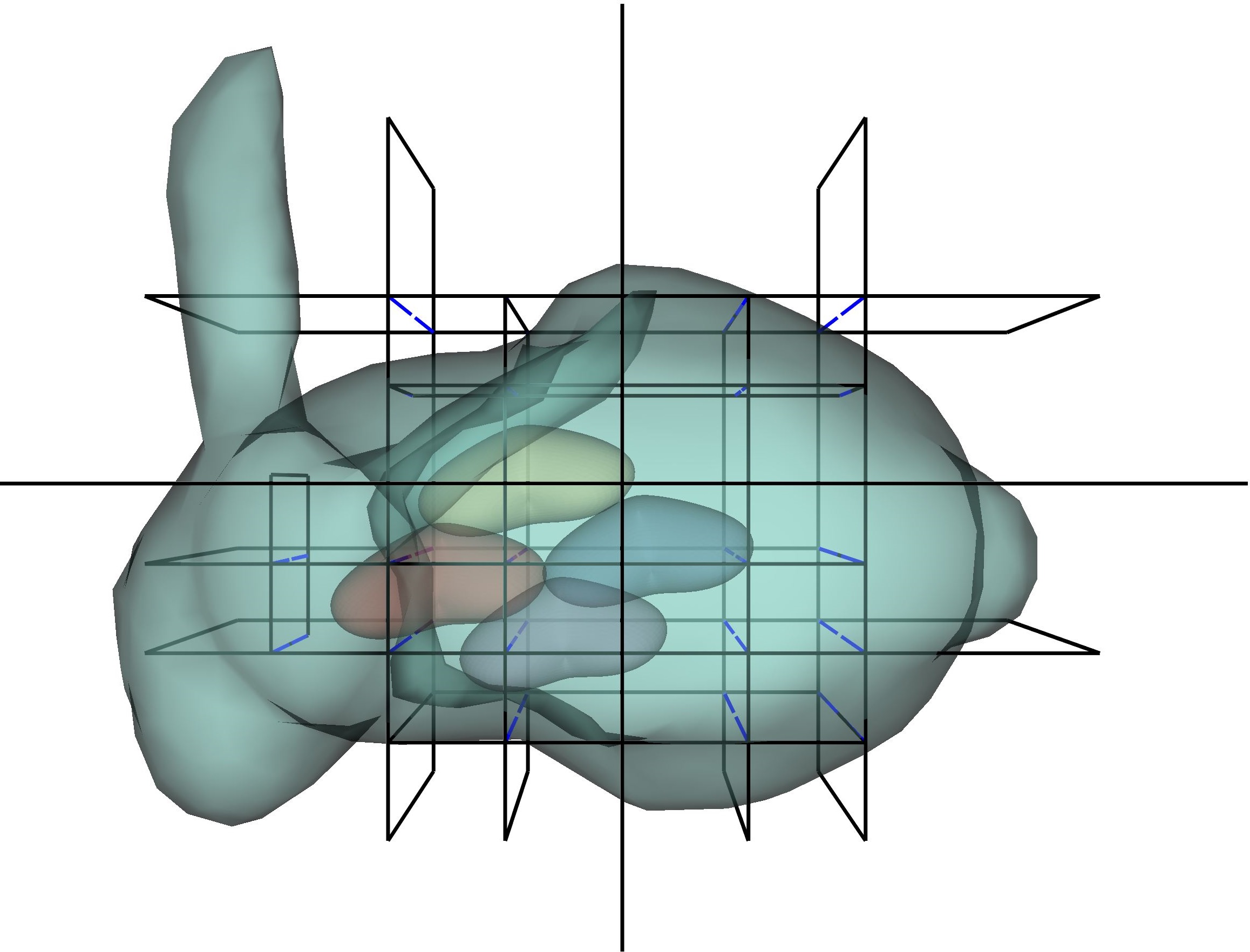} \\
   (b) $L=3$\\
  \includegraphics[height=1.2in]{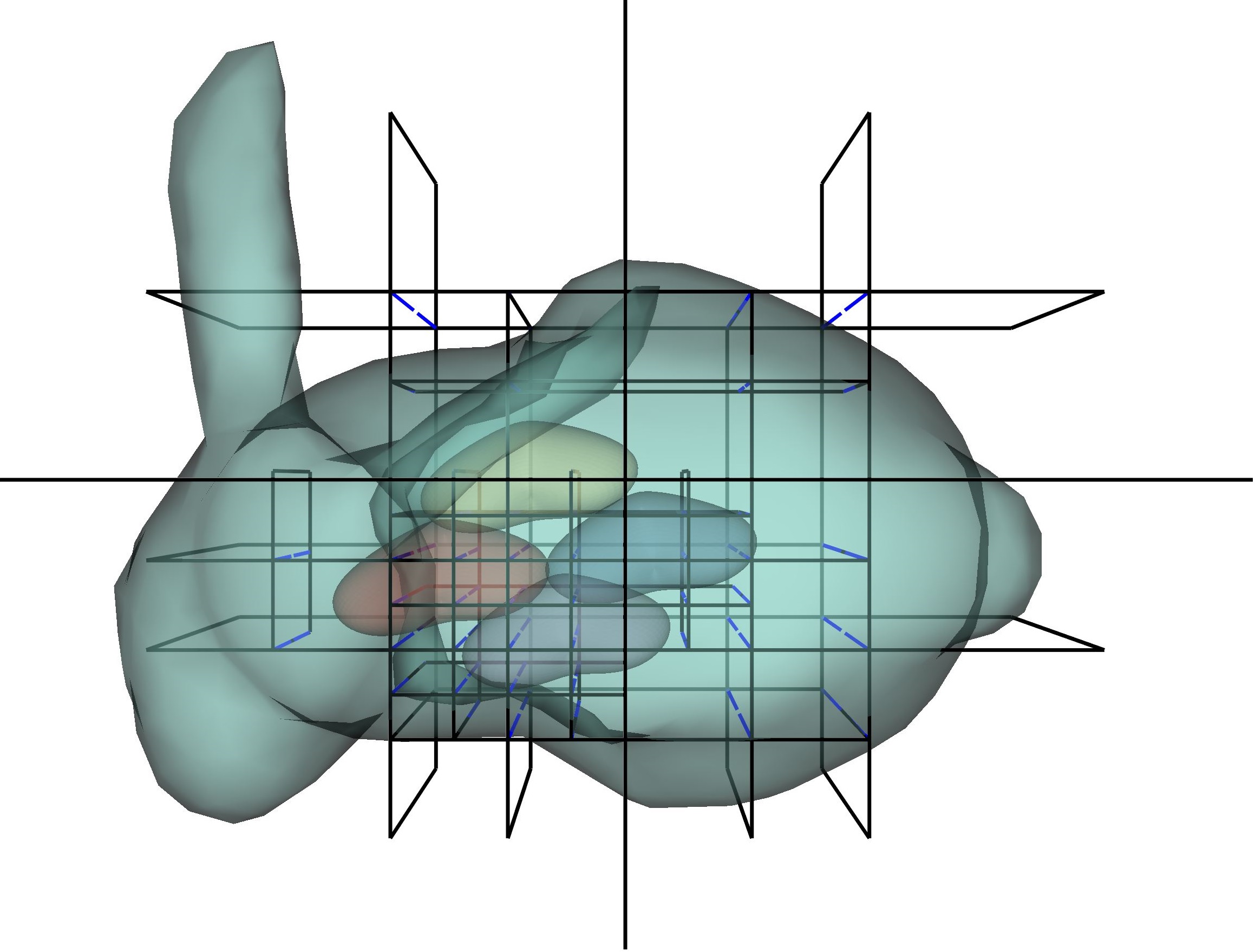} \\
   (c) $L=4$\\
 \end{tabular}
	}
	\caption{\XX{Results for different levels ($L$) of octree partitioning, for an artificial configuration of five meshes (top view).}}
	\label{fig:bunnies}
\end{minipage}
\vspace{-10pt}
\end{figure*}

\noindent{\textbf{Nested Spheres Dataset}}---We created a synthetic dataset consisting of four sphere meshes, nested in the 3D spatial configuration, shown in Figure~\ref{fig:result_mesh}(a).
The octree computation with $L=3$ yields twelve slices, the assembly order of which, and the single A3 paper placement is shown in Figure~\ref{fig:result_mesh}(b).
The resulting sliceform is shown in Figure~\ref{fig:result_mesh}(c). 
This example, despite being simple in terms of depicted structures, is quite complex, as it has six cut-throughs for the deeper-nested structures.
The configuration of the nested spheres is identifiable, and the inner red sphere is visible. 

\noindent{\textbf{Head Dataset}}---We obtained from the BodyParts3D database~\cite{mitsuhashi2009bodyparts3d} 87 meshes comprising the anatomy of the head, and including the skull bones and brain components, as shown in Figure~\ref{fig:result_mesh}(d). 
The octree computation with $L=2$ yields six slices, and with $L=3$ yields fourteen slices. 
The order of the highest level, and its multi-A3 paper placement is shown in Figure~\ref{fig:result_mesh}(e).
The resulting sliceforms are shown in Figure~\ref{fig:result_mesh}(f), for both octree levels. 
This example, despite consisting of many meshes, is quite easy to assemble. 
The $L=2$ sliceform is easier, as it consists of only intersections---thus, it is straightforward to assemble, even for smaller sized slices. 
The $L=3$ sliceform includes also two complex cut-throughs, which had to cross seven slices each. 
Therefore, it is advised to use bigger paper formats. 
The configuration of the different structures in the human head can be clearly seen, although in the $L=2$ sliceform, the structures might be easier to identify.

\noindent{\textbf{Heart Dataset}}---We obtained from the BodyParts3D database~\cite{mitsuhashi2009bodyparts3d} 125 meshes comprising the anatomy of the heart, as shown in Figure~\ref{fig:result_mesh}(g). 
The octree computation with $L=2$ yields six slices, and with $L=3$ yields fourteen slices. 
The order of the lowest level, and its single A3 paper placement is shown in Figure~\ref{fig:result_mesh}(h).
The resulting sliceform is shown in Figure~\ref{fig:result_mesh}(i), for $L=2$. 
This example consists only of intersecting slices, therefore, it is quite easy to assembly, for both levels. 
For $L=3$, it is advised to use bigger paper formats or multiple pages.  
The configuration of the different structures in the human heart is clearly represented, and the heart chambers are identifiable.

\XX{\noindent{\textbf{Stuffed Bunny Dataset}}---We created a synthetic dataset consisting of five meshes (Stanford bunny, and four hearts) in the 3D spatial configuration shown in Figure~\ref{fig:bunnies}.
This example includes meshes, more topologically complex than the nested spheres.
In Figure~\ref{fig:bunnies}, we showcase the effect of adjusting the parameter $L$ from $2$ to $4$, in the octree partitioning. 
Increasing the value of $L$ creates more slices in the denser sub-region, to reveal higher details of the structures.}

\begin{figure*}
\begin{minipage}{0.62\textwidth}
\centering{
  \setlength{\tabcolsep}{3pt}
 \begin{tabular}{ccc}
  \includegraphics[height=1.15in]{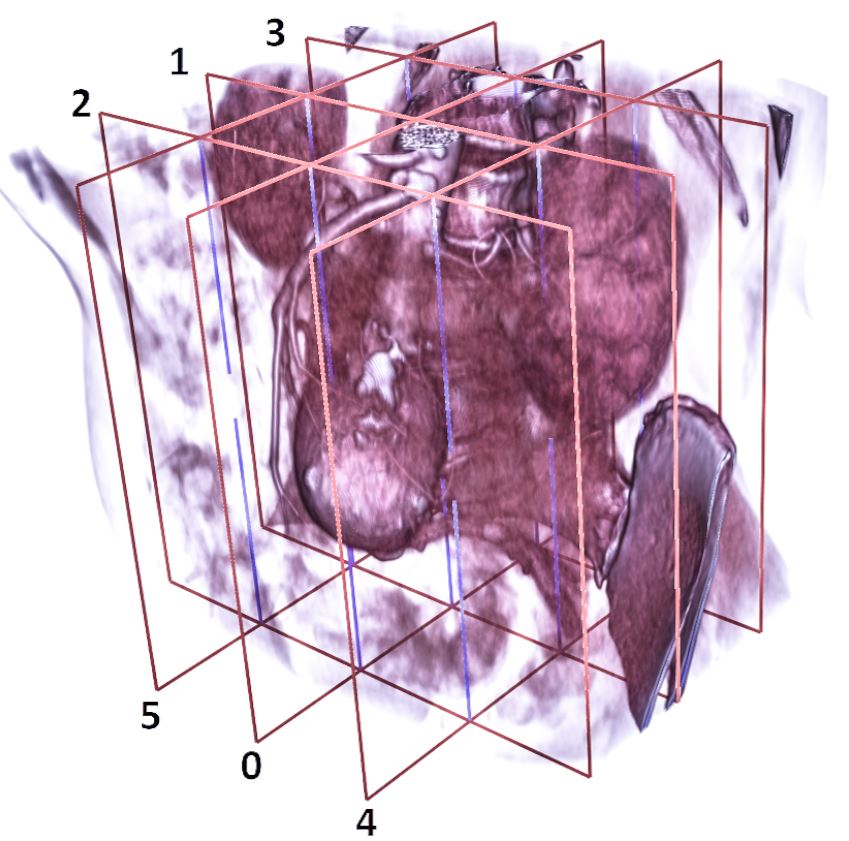} &
  \includegraphics[height=1.15in]{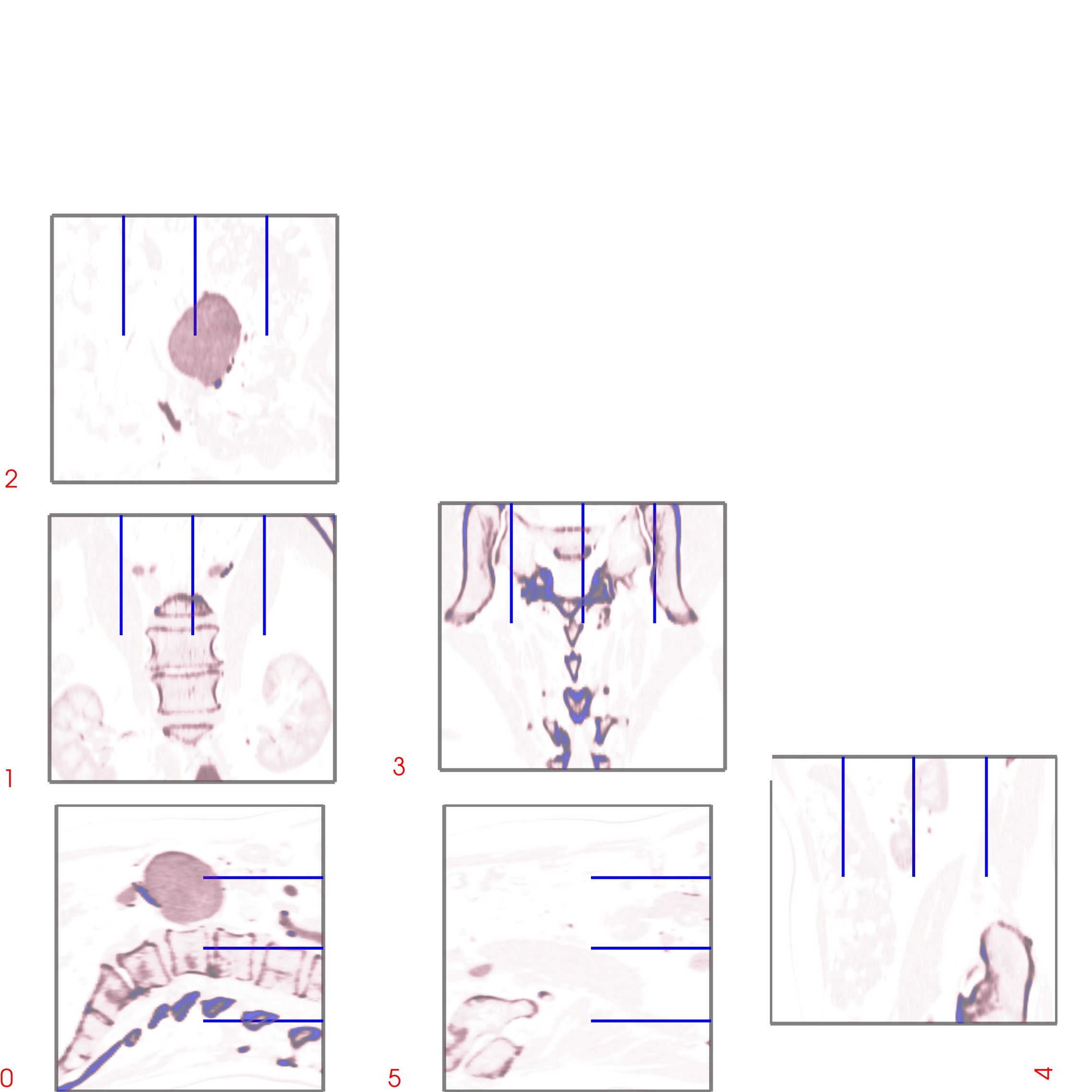} &
  \includegraphics[height=1.15in]{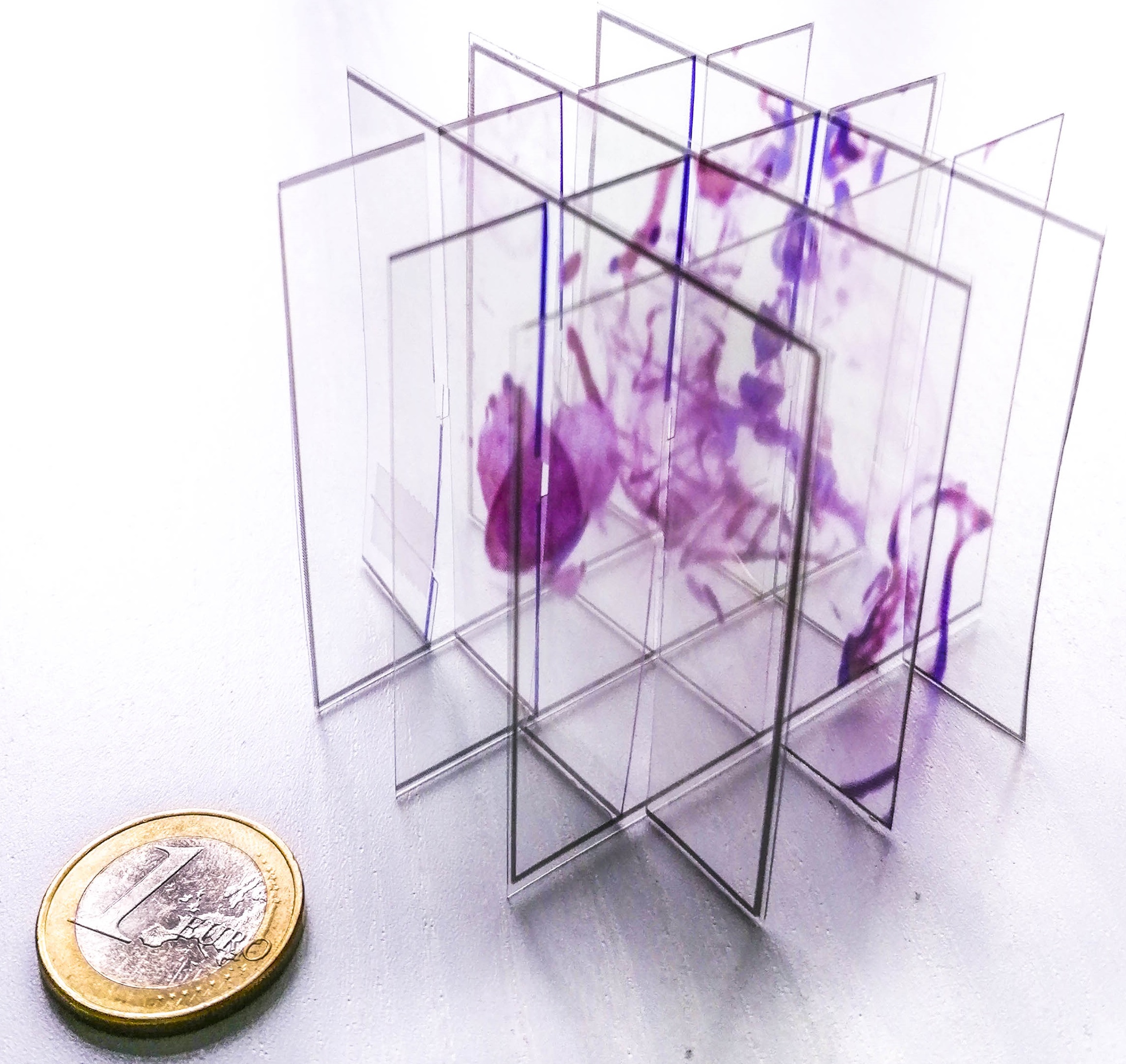} \\
   (a) & (b) & (c)\\
  \includegraphics[height=1.15in]{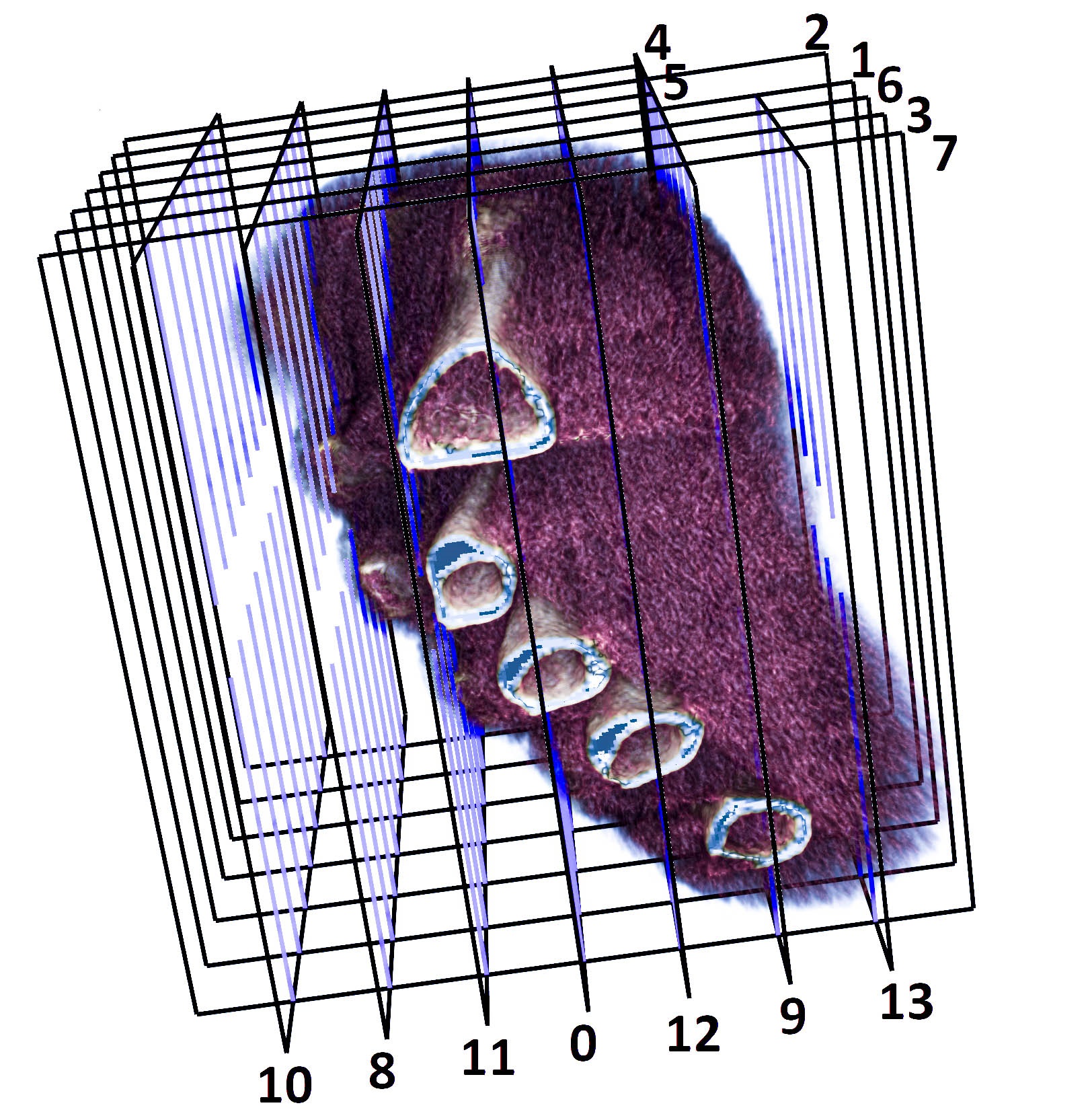} &
  \includegraphics[height=1.15in]{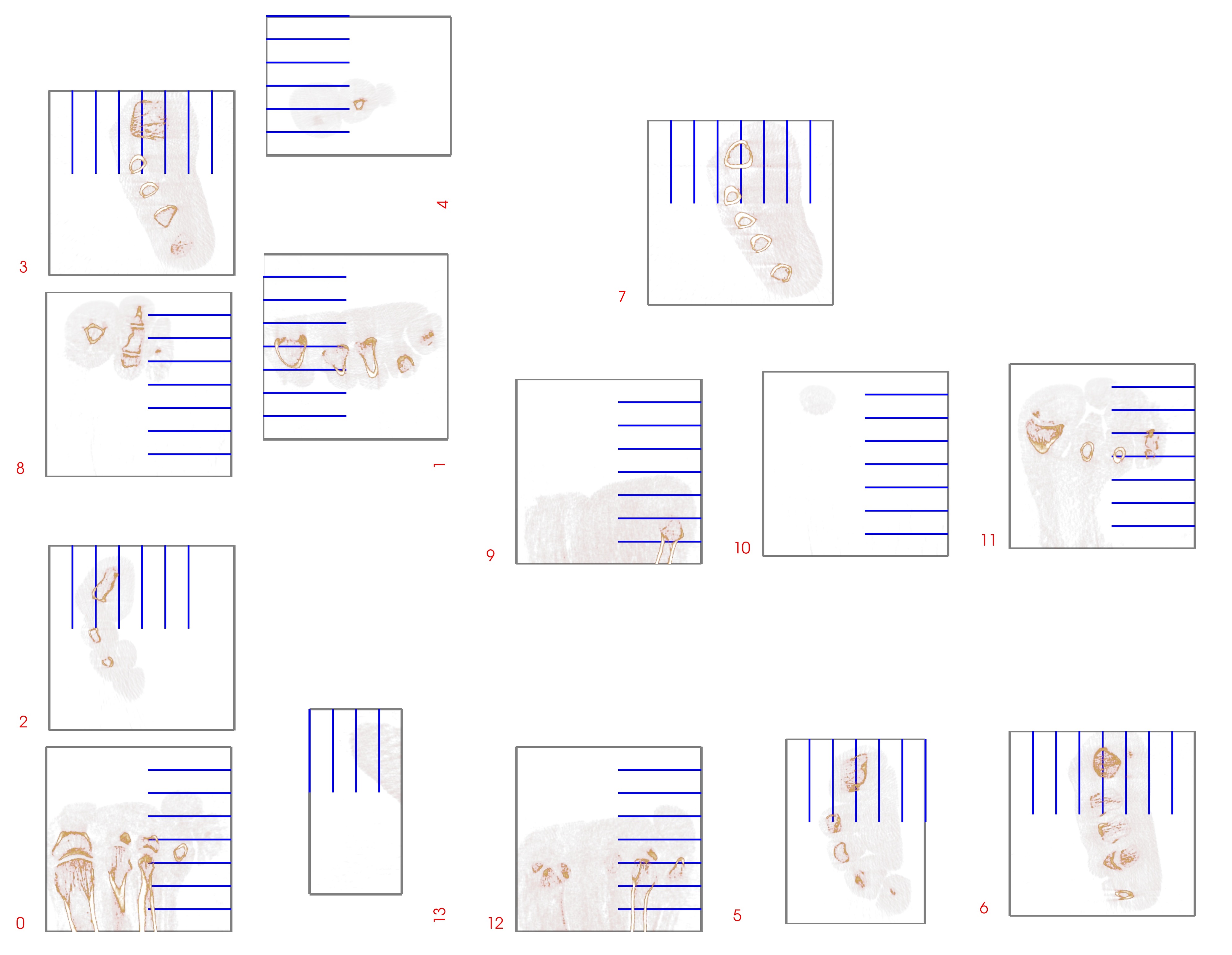} &
  \includegraphics[height=1.15in]{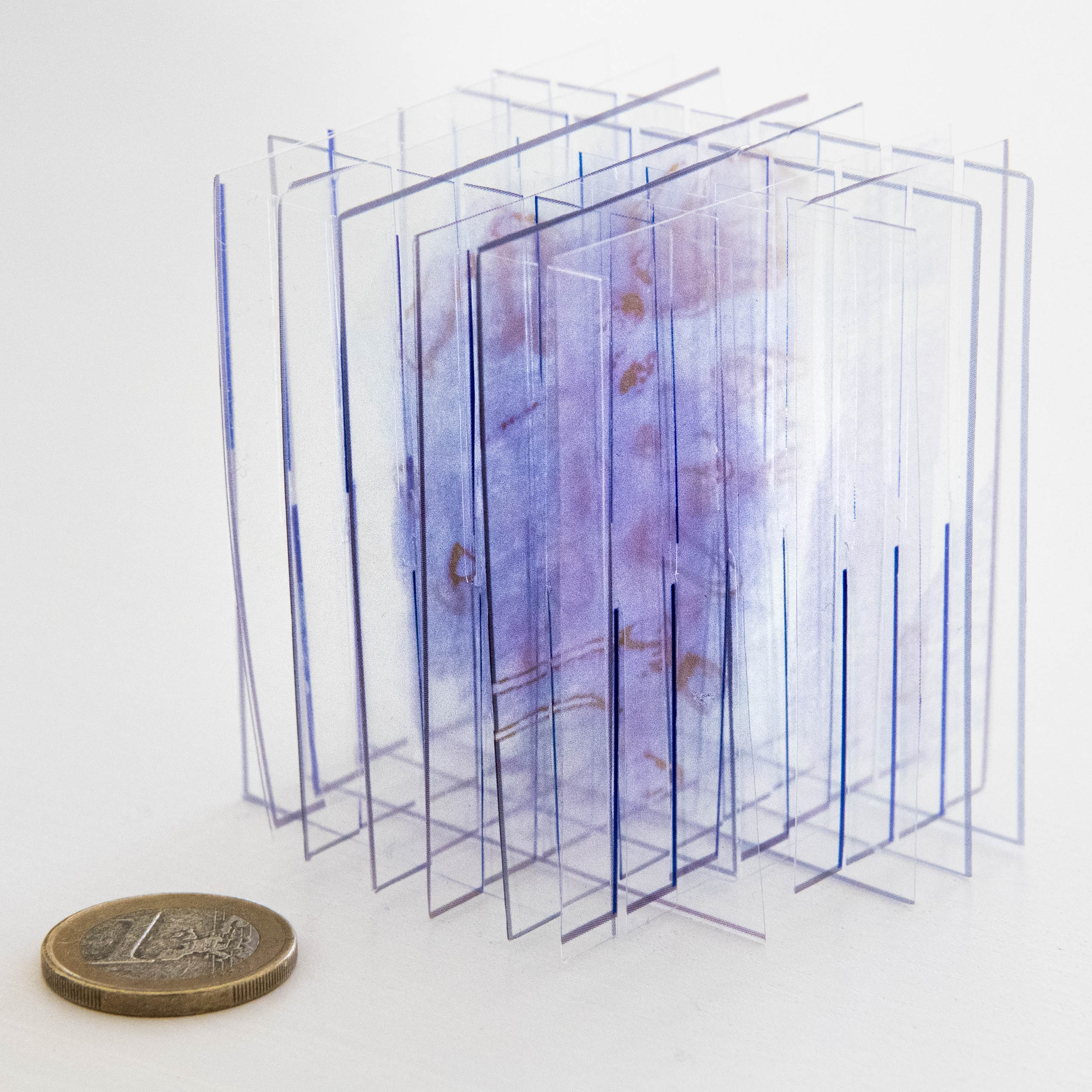} \\
  (d) & (e) & (f)\\
  \includegraphics[height=1.15in]{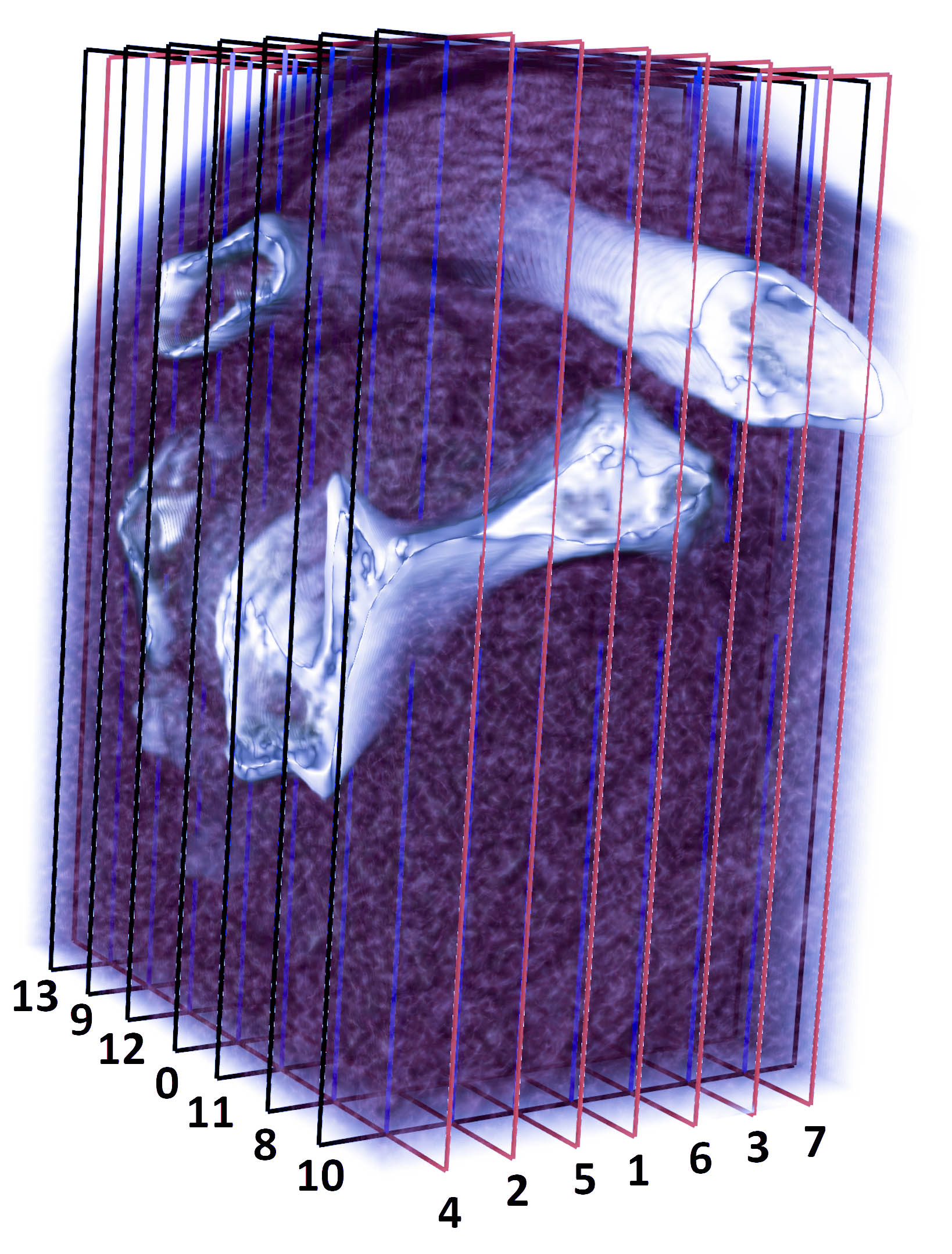} &
  \includegraphics[height=1.15in]{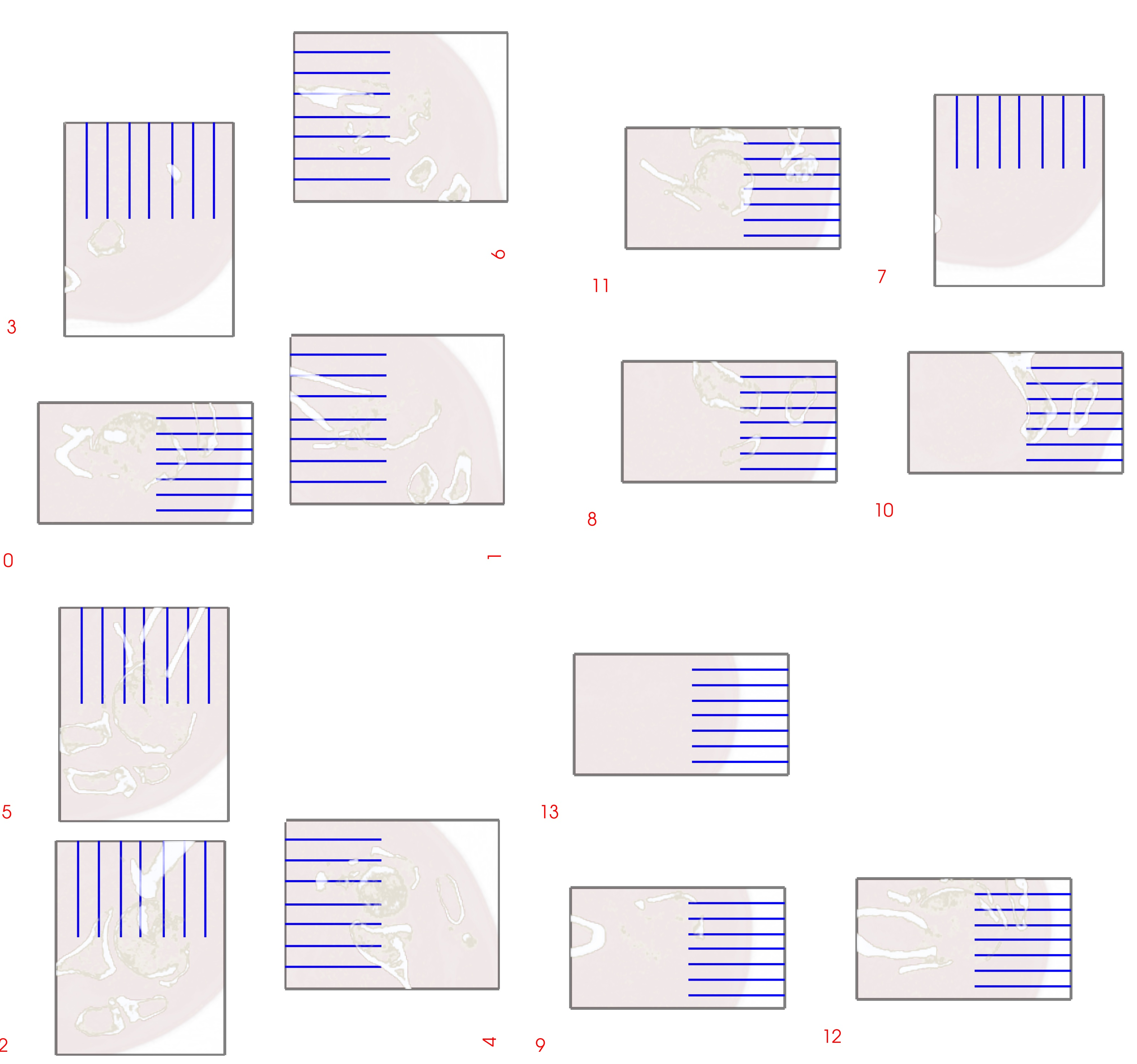} & \includegraphics[height=1.15in]{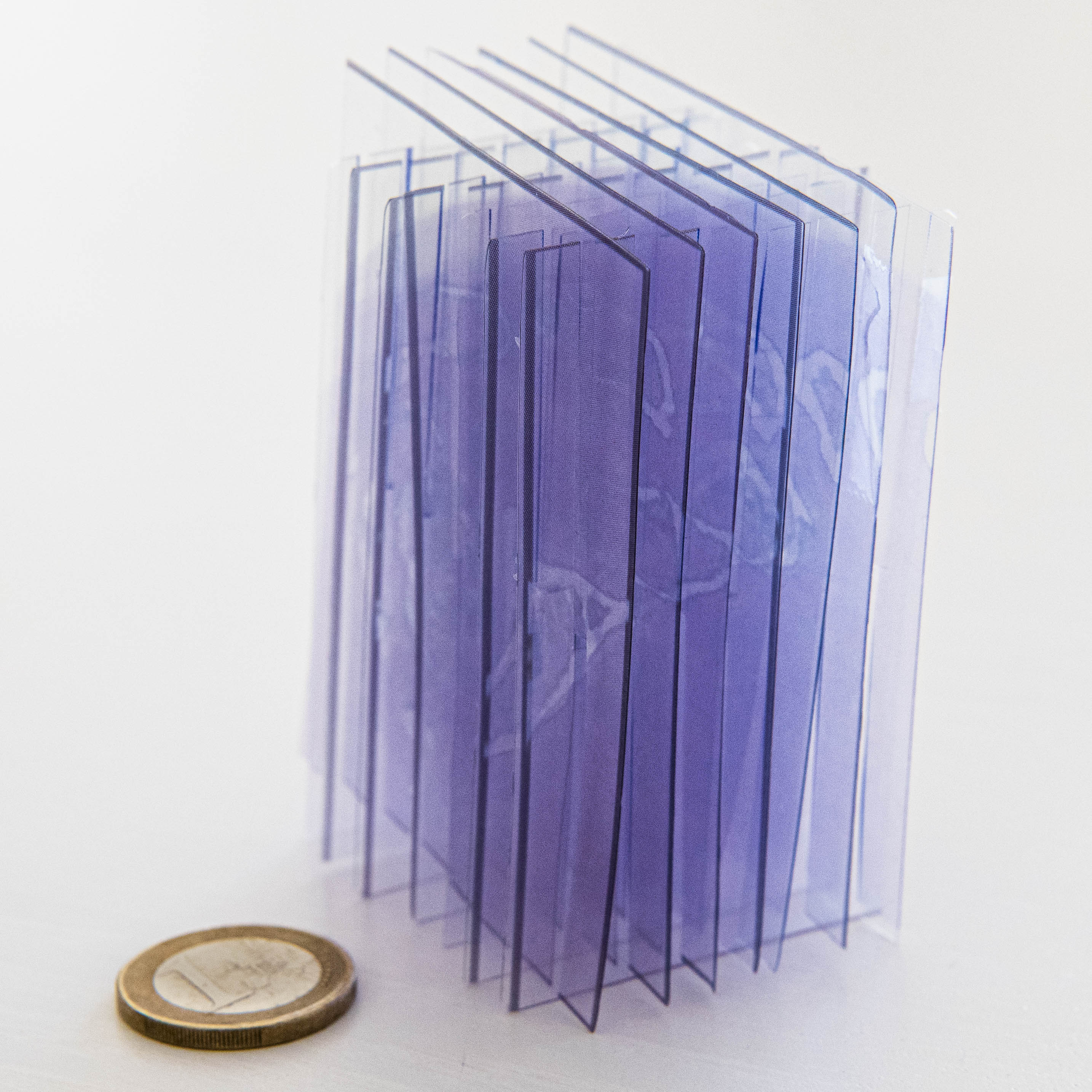}\\
  (g) & (h) & (i)\\
 \end{tabular}
}
\caption{Results for three CT volumes. In (a)--(c), we show a dataset of an aneurysm, in (d)--(f) of a foot, and in (g)--(i) of a shoulder.\vspace{-10pt}}
\label{fig:result_vol}
\end{minipage} 
\begin{minipage}{0.01\textwidth}
\parbox{0.01\textwidth}{}
\end{minipage} 
\begin{minipage}{0.35\textwidth}
	\centering{
	  \setlength{\tabcolsep}{1pt}
 \begin{tabular}{c}
  \includegraphics[height=1.2in]{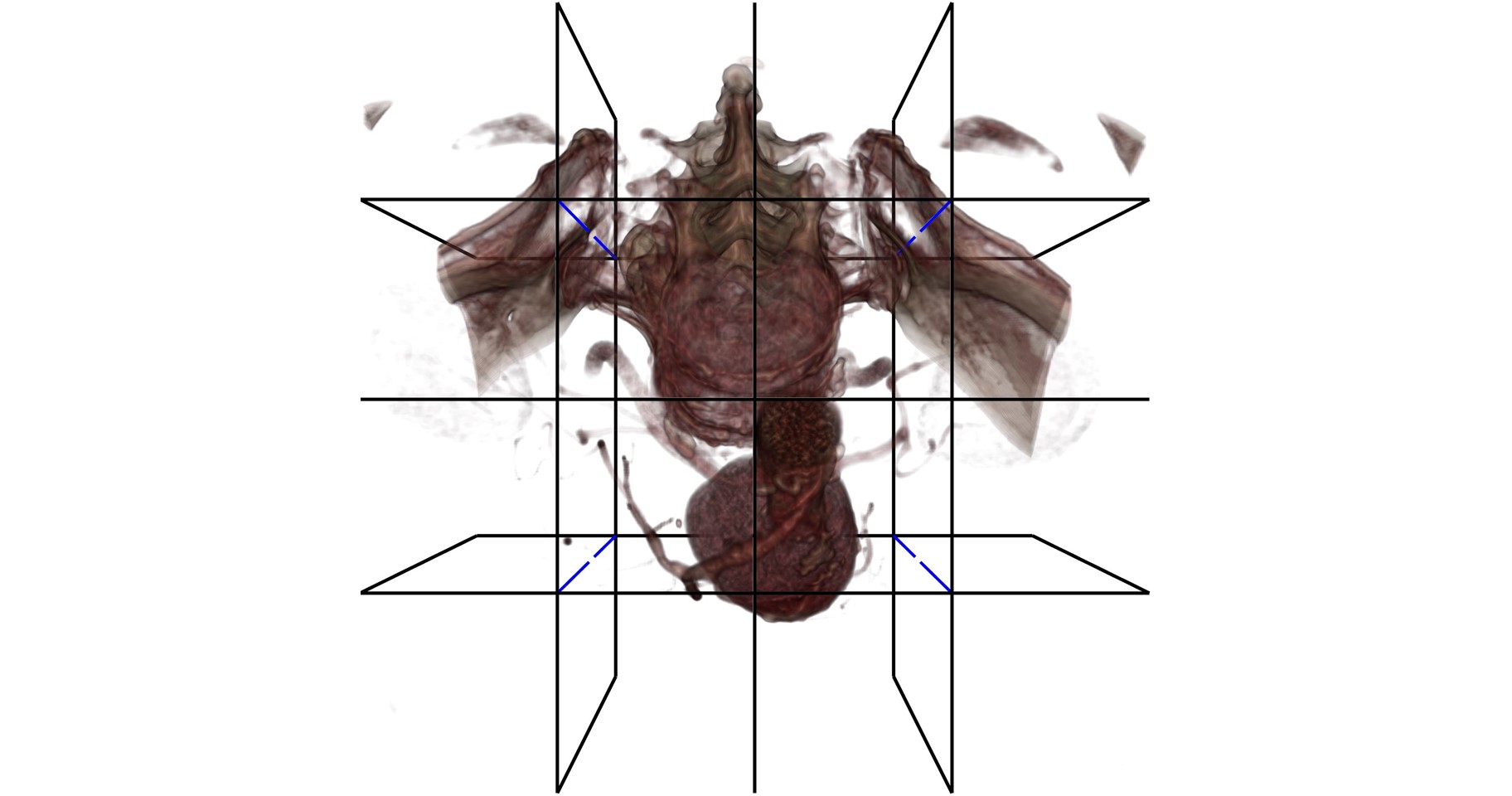} \\
   (a) $L=2$\\
  \includegraphics[height=1.2in]{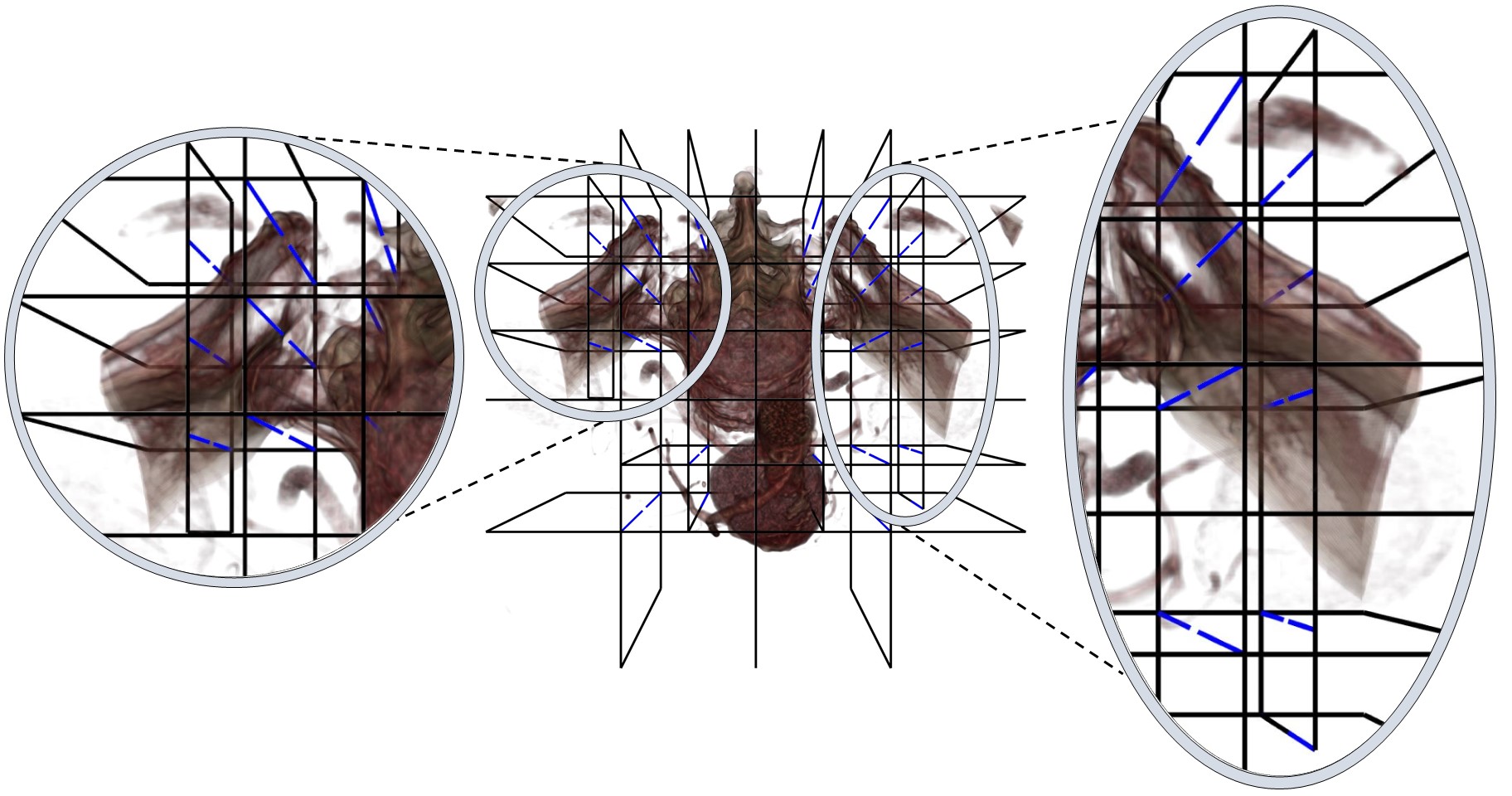} \\
   (b) $L=3$\\
  \includegraphics[height=1.2in]{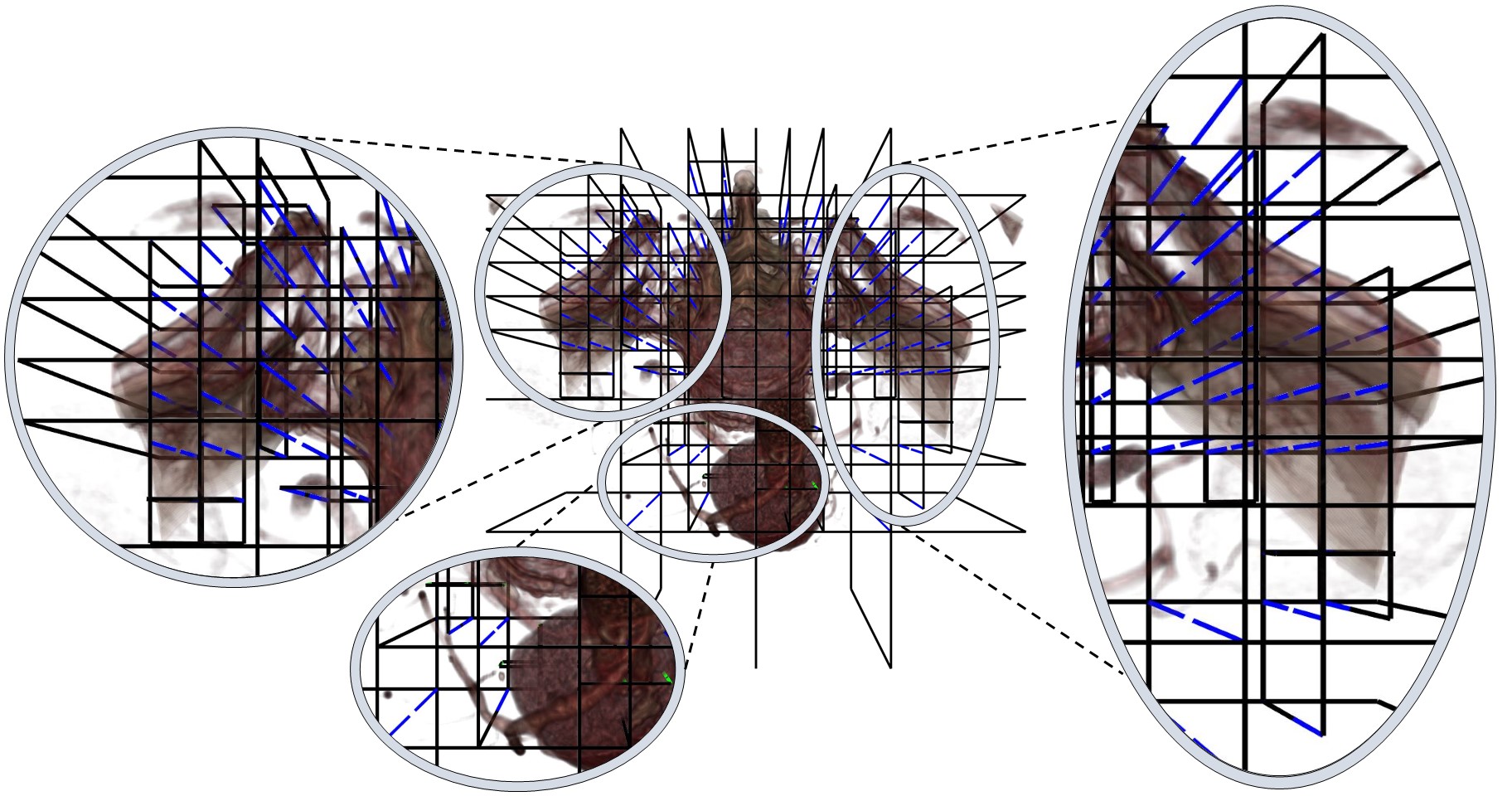} \\
   (c) $L=4$\\
 \end{tabular}
	}
	\caption{\XX{Results for different levels 
	($L$) of octree partitioning, for the aneurysm dataset (top view).}}
	\label{fig:levels_aneurysm}
\end{minipage}
\vspace{-10pt}
\end{figure*}

\section{Results with Volumetric Datasets}
\label{sec:vol_results}

\noindent{\textbf{Aneurysm Dataset}}---Our first volumetric example comes from a CT dataset of the lower torso of a patient with an aneurysm, as shown in Figure~\ref{fig:result_vol}(a). 
The spatial resolution of the dataset is $256 \times 257 \times 119$ voxels, with a spacing of $0.74 \times 0.74 \times 1.5~mm^3$.
The octree computation with $L=2$ yields six slices, which correspond to coronal and sagittal planes of the dataset. 
The order of assembly, and its single A3 paper placement are shown in Figure~\ref{fig:result_vol}(b).
The resulting sliceform is shown in Figure~\ref{fig:result_vol}(c). 
This example consists only of intersecting slices, therefore, it is quite easy to assembly, both in A4 and in A3 size.
The pathology (aneurysm) of the patient is clearly visible, and the sliceform gives a pseudo-3D appearance on the data, while at the same time it shows them in a slice-based manner, similar to clinical practice.
With an optimized transfer function, the view on the inner structure of the patient's body would also improve. 
\XX{Figure~\ref{fig:levels_aneurysm} shows this dataset with a different transfer function and at various levels ($L=2$ to $L=4$) of partitioning. 
As the octree level increases, additional slices provide a better representation of fine structures within the dataset, such as more details on the blood vessels, and on smaller bones.}

\noindent{\textbf{Foot Dataset}}---Our second volumetric example comes from a CT dataset of a human foot, as shown in Figure~\ref{fig:result_vol}(d). 
The spatial resolution of the dataset is $256 \times 256 \times 256$ voxels, with a spacing of $1 \times 1 \times 1~mm^3$.
The octree computation with $L=3$ yields fourteen slices, which correspond to axial and sagittal planes of the dataset. 
The order of assembly, and its single A3 paper placement are shown in Figure~\ref{fig:result_vol}(e).
The resulting sliceform is shown in Figure~\ref{fig:result_vol}(f). 
This example consists of ten intersecting slices, and four cut-throughs without stoppers, and is more complex than the previous one.
The sliceform gives a pseudo-3D appearance to the foot, similar to a hologram. 
The bones and other small structures are visible, and can be explored, but probably an optimized transfer function would be required for a better view on the volume. 

\noindent{\textbf{Shoulder Dataset}}---The last volumetric example comes from a CT dataset of a patient with a broken shoulder, as shown in Figure~\ref{fig:result_vol}(g). 
The spatial resolution of the dataset is $131 \times 94 \times 166$ voxels, with a spacing of $1 \times 1 \times 1~mm^3$.
The octree computation with $L=3$ yields fourteen slices, which correspond to axial and coronal planes of the dataset. 
The order of assembly, and its multi-A3 paper placement are shown in Figure~\ref{fig:result_vol}(h).
The resulting sliceform is shown in Figure~\ref{fig:result_vol}(i). 
This example consists only of intersecting slices, which are quite close to each other, therefore, it is quite difficult to assemble the sliceform, and also to discern clearly the bone structures.

\XX{\section{Evaluating the Feasibility of Sliceforms}}
\label{sec:evaluate}

We conducted an evaluation with $10$ participants, to determine the feasibility and time-efficiency of our proposed approach.
Our group of participants was gender-balanced (five males, five females), and they were between 28 and 37 years old.
At this initial stage, we decided not to include children, as our evaluation would require a different design than for adults. 
Two of our participants have normal vision, three were wearing lenses, and five were wearing glasses. 
Six participants are from the visualization community (one in BioVis, one in BioMedVis, one in MedVis, and the rest in InfoVis or Visual Analytics), one is a software engineer, one is a software developer, one is an account manager, and one is a teacher. 
One of the participants is quite keen on paper crafting, and puzzles.

We designed our evaluation based on guidelines from the seven scenarios by Lam et al.~\cite{ lam2011empirical}.
We were interested in evaluating three aspects: (i) the assembly performance of the users, (ii) the communicative value of our proposed physicalizations, and (iii) the experience of the users, while constructing the physicalizations.
To this end, we conducted a controlled experiment, where we asked our evaluation participants to assemble a set of sliceforms, while we measured the time for completion (\textit{User Performance}). 
At the same time, we observed their behavior and interaction with the sliceforms, during and after assembly (\textit{Communicative Value}). 
After the controlled experiment, we asked them open questions about their experience with the sliceform assembly, to obtain feedback (\textit{User Experience}). 

We prepared three cases, where slices were printed, and cut with the use of a cutting plotter, for precision.
A \emph{pop-out cutting} setting, which performs perforation cuts at slice boundaries and hinges, was selected to simulate the pre-cut sheets commonly used in papercraft books, to avoid that our users need to cut themselves, and to ensure the precision of the cuts.
The participants had only to take out the pre-cut slices from within the paper frame, and to assemble them. 
This time was not considered part of the experiment. 
Initially, we gave a short introduction and instructions on how the participants should interpret the annotations, and the labels on the printouts. 
We did not disclose to the participants the intended purpose of the sliceforms, but we showed them a printout of the 3D virtual counterpart of the dataset that they were building, and explained to them what the dataset is, and what it shows.
The participants could not interact with the 3D representation, as we wanted them to interact only with the physical object, and to determine whether they actually needed the interactive 3D representation. 
We asked our participants to construct the first two (aneurysm and spheres datasets, shown in the previous sections), while the third one (head mesh dataset, shown in Section~\ref{sec:mesh_results}) was optional, as it has more slices, is more complex and would definitely require more time, and more experience. 
We did not randomize our datasets, as they were intended to be shown in order of increasing difficulty. 
We did not train the participants beforehand, to test the feasibility of our designed papercraft.
Similarly, in a real-life scenario, they would start from an easy sliceform, and move on to harder cases. 
We measured the time for the completion of the assembly and upon completion, we stopped the time, inspected the sliceform, and informed them whether they had assembled it correctly, or not. 
If the assembly was wrong, we asked them to redo it, and measured the time again.

\begin{table*}[t]
\centering
\caption{Results of the user performance (UP) assessment, for evaluating the feasibility of the sliceforms.}
\label{tab:my-table}
\resizebox{.9\textwidth}{!}{%
\begin{tabular}{lccllrrrrrrrrr}
\cline{6-14}
 &  &  & \multicolumn{1}{c}{} & \multicolumn{1}{c|}{} & \multicolumn{3}{c|}{\textbf{Dataset 1: Aneurysm CT}} & \multicolumn{3}{c|}{\textbf{Dataset 2: Nested Spheres}} & \multicolumn{3}{c|}{\textbf{Dataset 3: Head Mesh Data}} \\ \cline{2-14} 
\multicolumn{1}{l|}{} & \multicolumn{1}{c|}{\textbf{Gender}} & \multicolumn{1}{c|}{\textbf{Age}} & \multicolumn{1}{c|}{\textbf{Vision}} & \multicolumn{1}{c|}{\textbf{Job}} & \multicolumn{1}{c|}{\textbf{Attempt 1}} & \multicolumn{1}{c|}{\textbf{Attempt 2}} & \multicolumn{1}{c|}{\textbf{Total attempts}} & \multicolumn{1}{c|}{\textbf{Attempt 1}} & \multicolumn{1}{c|}{\textbf{Attempt 2}} & \multicolumn{1}{c|}{\textbf{Total attempts}} & \multicolumn{1}{c|}{\textbf{Attempt 1}} & \multicolumn{1}{c|}{\textbf{Attempt 2}} & \multicolumn{1}{c|}{\textbf{Total attempts}} \\ \hline
\multicolumn{1}{|l|}{\textbf{A}} & \multicolumn{1}{c|}{F} & \multicolumn{1}{c|}{33} & \multicolumn{1}{l|}{lenses} & \multicolumn{1}{l|}{Account manager} & \multicolumn{1}{r|}{\textit{00:59}} & \multicolumn{1}{r|}{01:35} & \multicolumn{1}{r|}{2} & \multicolumn{1}{r|}{12:55} & \multicolumn{1}{r|}{-} & \multicolumn{1}{r|}{1} & \multicolumn{1}{r|}{34:00} & \multicolumn{1}{r|}{-} & \multicolumn{1}{r|}{1} \\ \hline
\multicolumn{1}{|l|}{\textbf{B}} & \multicolumn{1}{c|}{M} & \multicolumn{1}{c|}{28} & \multicolumn{1}{l|}{lenses} & \multicolumn{1}{l|}{BioVis} & \multicolumn{1}{r|}{01:41} & \multicolumn{1}{r|}{-} & \multicolumn{1}{r|}{1} & \multicolumn{1}{r|}{21:25} & \multicolumn{1}{r|}{-} & \multicolumn{1}{r|}{1} & \multicolumn{1}{r|}{-} & \multicolumn{1}{r|}{-} & \multicolumn{1}{r|}{-} \\ \hline
\multicolumn{1}{|l|}{\textbf{C}} & \multicolumn{1}{c|}{F} & \multicolumn{1}{c|}{28} & \multicolumn{1}{l|}{normal} & \multicolumn{1}{l|}{Software Developer} & \multicolumn{1}{r|}{\textit{02:40}} & \multicolumn{1}{r|}{04:44} & \multicolumn{1}{r|}{2} & \multicolumn{1}{r|}{\textit{15:30}} & \multicolumn{1}{r|}{29:07} & \multicolumn{1}{r|}{2} & \multicolumn{1}{r|}{-} & \multicolumn{1}{r|}{-} & \multicolumn{1}{r|}{-} \\ \hline
\multicolumn{1}{|l|}{\textbf{D}} & \multicolumn{1}{c|}{M} & \multicolumn{1}{c|}{33} & \multicolumn{1}{l|}{glasses} & \multicolumn{1}{l|}{BioMedVis} & \multicolumn{1}{r|}{\textit{01:16}} & \multicolumn{1}{r|}{02:31} & \multicolumn{1}{r|}{2} & \multicolumn{1}{r|}{18:59} & \multicolumn{1}{r|}{-} & \multicolumn{1}{r|}{1} & \multicolumn{1}{r|}{-} & \multicolumn{1}{r|}{-} & \multicolumn{1}{r|}{-} \\ \hline
\multicolumn{1}{|l|}{\textbf{E}} & \multicolumn{1}{c|}{M} & \multicolumn{1}{c|}{36} & \multicolumn{1}{l|}{lenses} & \multicolumn{1}{l|}{Visual Analytics} & \multicolumn{1}{r|}{01:15} & \multicolumn{1}{r|}{-} & \multicolumn{1}{r|}{1} & \multicolumn{1}{r|}{15:00} & \multicolumn{1}{r|}{-} & \multicolumn{1}{r|}{1} & \multicolumn{1}{r|}{-} & \multicolumn{1}{r|}{-} & \multicolumn{1}{r|}{-} \\ \hline
\multicolumn{1}{|l|}{\textbf{F}} & \multicolumn{1}{c|}{M} & \multicolumn{1}{c|}{33} & \multicolumn{1}{l|}{normal} & \multicolumn{1}{l|}{Software Engineer} & \multicolumn{1}{r|}{\textit{01:48}} & \multicolumn{1}{r|}{03:18} & \multicolumn{1}{r|}{2} & \multicolumn{1}{r|}{08:17} & \multicolumn{1}{r|}{-} & \multicolumn{1}{r|}{1} & \multicolumn{1}{r|}{23:33} & \multicolumn{1}{r|}{-} & \multicolumn{1}{r|}{1} \\ \hline
\multicolumn{1}{|l|}{\textbf{G}} & \multicolumn{1}{c|}{M} & \multicolumn{1}{c|}{29} & \multicolumn{1}{l|}{glasses} & \multicolumn{1}{l|}{Visual Analytics} & \multicolumn{1}{r|}{01:29} & \multicolumn{1}{r|}{-} & \multicolumn{1}{r|}{1} & \multicolumn{1}{r|}{14:34} & \multicolumn{1}{r|}{-} & \multicolumn{1}{r|}{1} & \multicolumn{1}{r|}{25:15} & \multicolumn{1}{r|}{-} & \multicolumn{1}{r|}{1} \\ \hline
\multicolumn{1}{|l|}{\textbf{H}} & \multicolumn{1}{c|}{F} & \multicolumn{1}{c|}{33} & \multicolumn{1}{l|}{glasses} & \multicolumn{1}{l|}{MedVis} & \multicolumn{1}{r|}{01:24} & \multicolumn{1}{r|}{-} & \multicolumn{1}{r|}{1} & \multicolumn{1}{r|}{08:32} & \multicolumn{1}{r|}{-} & \multicolumn{1}{r|}{1} & \multicolumn{1}{r|}{24:09} & \multicolumn{1}{r|}{-} & \multicolumn{1}{r|}{1} \\ \hline
\multicolumn{1}{|l|}{\textbf{I}} & \multicolumn{1}{c|}{F} & \multicolumn{1}{c|}{37} & \multicolumn{1}{l|}{glasses} & \multicolumn{1}{l|}{InfoVis} & \multicolumn{1}{r|}{02:35} & \multicolumn{1}{r|}{-} & \multicolumn{1}{r|}{1} & \multicolumn{1}{r|}{08:27} & \multicolumn{1}{r|}{-} & \multicolumn{1}{r|}{1} & \multicolumn{1}{r|}{26:38} & \multicolumn{1}{r|}{-} & \multicolumn{1}{r|}{1} \\ \hline
\multicolumn{1}{|l|}{\textbf{J}} & \multicolumn{1}{c|}{F} & \multicolumn{1}{c|}{30} & \multicolumn{1}{l|}{glasses} & \multicolumn{1}{l|}{Teacher} & \multicolumn{1}{r|}{02:14} & \multicolumn{1}{r|}{-} & \multicolumn{1}{r|}{1} & \multicolumn{1}{r|}{10:51} & \multicolumn{1}{r|}{-} & \multicolumn{1}{r|}{1} & \multicolumn{1}{r|}{29:29} & \multicolumn{1}{r|}{-} & \multicolumn{1}{r|}{1} \\ \hline
\textit{} & \textit{} & \textit{} & \multicolumn{1}{c}{\textit{}} & \multicolumn{1}{c}{\textit{}} & \multicolumn{1}{l}{} & \multicolumn{1}{l}{} & \multicolumn{1}{l}{} & \multicolumn{1}{l}{} & \multicolumn{1}{l}{} &  & \multicolumn{1}{l}{} & \multicolumn{1}{l}{} &  \\ \cline{6-14} 
 &  &  &  & \multicolumn{1}{l|}{} & \multicolumn{2}{c|}{\textbf{Time (mm:ss)}} & \multicolumn{1}{c|}{\textbf{Attempts (\textit{N})}} & \multicolumn{2}{c|}{\textbf{Time (mm:ss)}} & \multicolumn{1}{c|}{\textbf{Attempts (\textit{N})}} & \multicolumn{2}{c|}{\textbf{Time (mm:ss)}} & \multicolumn{1}{c|}{\textbf{Attempts (\textit{N})}} \\ \hline
\multicolumn{5}{|c|}{\textbf{Average ($\mu$)}} & \multicolumn{2}{r|}{02:17} & \multicolumn{1}{r|}{1.40} & \multicolumn{2}{r|}{14:45} & \multicolumn{1}{r|}{1.10} & \multicolumn{2}{r|}{27:11} & \multicolumn{1}{r|}{1} \\ \hline
\multicolumn{5}{|c|}{\textbf{Standard deviation ($\sigma$)}} & \multicolumn{2}{r|}{01:05} & \multicolumn{1}{r|}{0.52} & \multicolumn{2}{r|}{06:47} & \multicolumn{1}{r|}{0.32} & \multicolumn{2}{r|}{03:57} & \multicolumn{1}{r|}{0} \\ \hline
\end{tabular}%
}
\end{table*}

\noindent \textbf{User Performance (UP)}---For assessing UP, we measured the assembly completion time.
All participants constructed the aneurysm and the spheres dataset successfully. 
Only six participants volunteered to do the third experiment, as they were motivated by the previous experiments.
The completion times are summarized in Table~\ref{tab:my-table}, as well as the number of attempts. 
The average time ($mm:ss$) for the completion of the sliceforms was 02:17 ($\sigma=$01:05) for the aneurysm dataset, 14:45 ($\sigma=$06:47) for the spheres dataset and 27:11 ($\sigma=$03:57) for the head mesh dataset. 
For the completion of the aneurysm sliceform, six people finished it with one attempt, and the others needed a second attempt. 
For the spheres dataset, only one needed a second attempt, and for the head mesh dataset, all six participants completed it with one attempt. 
This could be due to learning, from the previous simpler cases. 
Although the fastest participant has a MedVis background, we did not observe any statistically significant difference in performance times between participants with a visualization (or Bio/MedVis) background, and the people that had no prior knowledge. 
Also, we did not observe any difference between men and women.
All participants were able to recognize that they had committed an error in the assembly, and we only confirmed that, so that they could start the second attempt. 
We cannot quantify whether the second attempt needed more or less time, as participants committed errors at different steps of the assembly, and some needed to restart from scratch, while others from just half-way through.

\noindent \textbf{Communication Value (CV)}---For assessing CV, we observed how the participants were assembling, interacting, and exploring the sliceform, during the previously described controlled experiment.
Some participants were using the order of assembly and others not, since we did not force them to follow the instructions, but kept this as a recommendation. 
The fastest participant used strictly the order for all datasets. 
The second fastest, though, used the order only for the last case. 
He did not use the order in the other two cases, because he thought that starting from the corners was easier than from the center, but in the interview, he commented that he should have used the order. 
The third, fourth, and fifth fastest participants strictly used the order, in all cases.
Two of the participants never used the order, and they had the most difficulties and the longest completion times.
We also noticed during this phase that most of the participants were really motivated to finish the sliceforms. 
Even if they found the assembly difficult, and were frustrated, none of them gave up.
In some cases, if the sliceforms were too small, and if the slices were too close together, they had difficulties with the assembly, and were slower, but started improvising and using objects to push the slices into the appropriate position. 
They all always wanted to see the 3D model with the slice placement, at all times. 
Most considered the aneurysm sliceform very easy, even those, who got it wrong.
This is probably due to having no prior experience with sliceforms, or the underlying data.
The spheres sliceform was considered tougher, but most of the people got it correct with the first attempt.
Most of the participants, after building the sliceform, started looking at it from all possible sides, to see if they built it correctly, and see how it looked like.
All were able to tell if they made a mistake by themselves.
Most showed satisfaction with their successful assembly of the sculpture, and some wanted to keep it. 

\XX{\noindent \textbf{User Experience (UE)}---After the experiment, we asked each participant a set of open questions. 
When asking whether sliceforms are helpful in explaining and communicating how the data looks in reality, and if they could extract useful information from building up, interacting, and observing the sliceforms, a participant (account manager) answered that \textit{``it gives a more 3D view on the data''}, and another participant (visualization researcher) that \textit{``It helps you build a mental image of the slice configuration. 
It looks like you are interpolating the space mentally and build a 3D view in your head and imagine how the entire space looks like.''} 
When asking about interaction and assembly experience, many of the participants commented that they wanted to finish it, even if it was difficult, and/or frustrating. 
This shows that sliceforms are actually engaging. 
About interaction after the assembly, one participant stated that
\textit{``during the assembly, I was focused only on the building and did not look at the data, but at the end, I realized that I had gotten a totally new view on the data.''}
With regard to missing features, points for improvements, and current limitations, the participants mentioned that sometimes the cuts are difficult to access, and that a stiffer material could be easier to handle. 
Many participants noticed that the complexity increases fast with an increasing number of slices, and with more complex configurations, in case of cut-throughs.
One important point for improvement is that the labels for the order should actually be printed on the slices, maybe at a corner, instead next to the slices, on the frame. 
One participant would have liked \textit{``a step-by-step IKEA-like instruction''}.
Two participants would have liked to have some visual indicators showing the three orientations (both visualization researchers). 
All participants thought that the assembly process was understandable, \textit{``but it needs practice and patience.''} 
With regard to the material, the participants said that using transparent films is \textit{``even more useful''}, because \textit{``when you see the intersections, it looks more spatial''}, and because \textit{``with paper, you cannot see through it. You need the transparent film to make a hologram out of it.''}
When asked how they would use sliceforms, people with kids commented that they would use it with them, \textit{``as a recreational game''}, or \textit{``as an educational toy''.} 
Others commented that it is \textit{``good to show the intrinsic structure of medical data''},  \textit{``a great example for teaching 3D rendering, to explain it better and easier''} (visualization researcher), \textit{``[good] for education or for explaining concepts, like an octree''} (software engineer).}

\vspace{8pt}
\XX{\section{Evaluating the Educational Value of Sliceforms}
We conducted an additional evaluation with $7$ participants, to determine the educational value of our approach for the general public.
Our group of participants included four males, and three females, between 26 and 39 years old.
Similarly to the feasibility evaluation above, we did not include children.
Three of our participants have normal vision, two were wearing lenses, and two were wearing glasses. 
The educational and occupational background of our participants is broad, and only one person has good knowledge of anatomy---being a nurse. 
\XX{\begin{table}[t]
\centering
\caption{Results of the educational value assessment.}
\label{tab:my-table2}
\resizebox{0.45\textwidth}{!}{%
\begin{tabular}{cccllccc}
\cline{2-8}
\multicolumn{1}{c|}{} & \multicolumn{1}{c|}{\textbf{Gender}} & \multicolumn{1}{c|}{\textbf{Age}} & \multicolumn{1}{c|}{\textbf{Vision}} & \multicolumn{1}{c|}{\textbf{Job}} & \multicolumn{1}{c|}{\textbf{\begin{tabular}[c]{@{}c@{}}Correct\\ Answers\end{tabular}}} & \multicolumn{1}{c|}{\textbf{\begin{tabular}[c]{@{}c@{}}Incorrect\\ Answers\end{tabular}}} & \multicolumn{1}{c|}{\textbf{Doubting}} \\ \hline
\multicolumn{1}{|c|}{\textbf{A}} & \multicolumn{1}{c|}{F} & \multicolumn{1}{c|}{33} & \multicolumn{1}{l|}{lenses} & \multicolumn{1}{l|}{Journalist} & \multicolumn{1}{c|}{9} & \multicolumn{1}{c|}{0} & \multicolumn{1}{c|}{1} \\ \hline
\multicolumn{1}{|c|}{\textbf{B}} & \multicolumn{1}{c|}{M} & \multicolumn{1}{c|}{35} & \multicolumn{1}{l|}{glasses} & \multicolumn{1}{l|}{Programmer} & \multicolumn{1}{c|}{8} & \multicolumn{1}{c|}{0} & \multicolumn{1}{c|}{2} \\ \hline
\multicolumn{1}{|c|}{\textbf{C}} & \multicolumn{1}{c|}{M} & \multicolumn{1}{c|}{39} & \multicolumn{1}{l|}{normal} & \multicolumn{1}{l|}{Business analyst} & \multicolumn{1}{c|}{9} & \multicolumn{1}{c|}{0} & \multicolumn{1}{c|}{1} \\ \hline
\multicolumn{1}{|c|}{\textbf{D}} & \multicolumn{1}{c|}{M} & \multicolumn{1}{c|}{31} & \multicolumn{1}{l|}{glasses} & \multicolumn{1}{l|}{Programmer} & \multicolumn{1}{c|}{6} & \multicolumn{1}{c|}{2} & \multicolumn{1}{c|}{2} \\ \hline
\multicolumn{1}{|c|}{\textbf{E}} & \multicolumn{1}{c|}{F} & \multicolumn{1}{c|}{26} & \multicolumn{1}{l|}{normal} & \multicolumn{1}{l|}{French teacher} & \multicolumn{1}{c|}{5} & \multicolumn{1}{c|}{3} & \multicolumn{1}{c|}{2} \\ \hline
\multicolumn{1}{|c|}{\textbf{F}} & \multicolumn{1}{c|}{M} & \multicolumn{1}{c|}{29} & \multicolumn{1}{l|}{normal} & \multicolumn{1}{l|}{Photographer} & \multicolumn{1}{c|}{8} & \multicolumn{1}{c|}{0} & \multicolumn{1}{c|}{2} \\ \hline
\multicolumn{1}{|c|}{\textbf{G}} & \multicolumn{1}{c|}{F} & \multicolumn{1}{c|}{30} & \multicolumn{1}{l|}{lenses} & \multicolumn{1}{l|}{Nurse} & \multicolumn{1}{c|}{10} & \multicolumn{1}{c|}{0} & \multicolumn{1}{c|}{0} \\ \hline
 &  &  & \multicolumn{1}{c}{} & \multicolumn{1}{c}{} &  &  &  \\ \hline
\multicolumn{5}{|c|}{\textbf{Average ($\mu$)}} & \multicolumn{1}{c|}{\cellcolor[HTML]{FFFFFF}7.86} & \multicolumn{1}{c|}{\cellcolor[HTML]{FFFFFF}0.57} & \multicolumn{1}{c|}{\cellcolor[HTML]{FFFFFF}1.43} \\ \hline
\multicolumn{5}{|c|}{\textbf{Standard deviation ($\sigma$)}} & \multicolumn{1}{c|}{\cellcolor[HTML]{FFFFFF}1.77} & \multicolumn{1}{c|}{\cellcolor[HTML]{FFFFFF}0.98} & \multicolumn{1}{c|}{\cellcolor[HTML]{FFFFFF}0.79} \\ \hline
\end{tabular}%
}
\end{table}}

We conducted a controlled experiment with three cases (aneurysm, head, and heart sliceforms), where we asked our evaluation participants to identify several anatomical and/or pathological structures in pre-assembled sliceforms. 
These structures were shown to them using traditional, static, labeled anatomical illustrations, retrieved online, and printed prior to the evaluation.
The purpose of this experiment was to evaluate whether anatomical structures can be identified and recognized in the sliceforms.
During the experiment, we randomized the order of datasets and structures. 
We did not measure the time that was needed to identify a structure, only whether it was successfully pointed out (identified, not identified, and doubting: correct, but unsure answer).
These measurements are summarized in Table~\ref{tab:my-table2}.
Most of the people could identify the anatomical structures correctly, or with some doubts. 
The nurse participant identified everything correctly.
The missed structures were either the aneurysm (a pathology that is not realistically depicted in an illustration), or structures that are deeper in the volume, such as the corpus callosum in the brain, or structures that are more difficult to tell apart from others, such as the right vs. left atrium in the heart.
In general, the orientation (left--right) was an issue for most participants.
We did not observe any difference between men and women.

After the controlled experiment, we asked them open questions about their learning experience with the sliceforms. 
The first question was whether there is something that they could see better or easier, either in the sliceform, or in the illustration. 
Most of the participants commented that they appreciate the three-dimensional view of the organs, as well as the ability to look at the structures from different viewpoints. 
One participant commented that it is ``\textit{easier to see the shape and the size of things}''. 
One person commented that ``\textit{the colors confuse me a little bit in model}'', and another one that ``\textit{sometimes, I cannot see very deep into the model, but maybe I can take out a few slides}''.
The nurse commented that the sliceform would be ``\textit{a nice toy for kids}'', and that she cannot see something that she did not already know. 
Regarding the effect of the physical model on learning anatomy compared to a simple anatomical illustration, five participants commented that it is more fun, and more interesting, being interactive. 
Two commented that it was easy, but they would have appreciated labels (``\textit{If I would have the names of the structures on the model, I could learn a lot---and faster}''). 
One participant commented that ``\textit{if it would take me a long time to make [the sliceform], maybe I would still prefer to google [an illustration]}''.
When we asked whether the tool is helpful in explaining and communicating how anatomy looks in reality, all participants (except for the nurse) answered positively. 
The nurse commented that she might use it for explaining anatomy to a patient, while another participant commented that ``\textit{it would be cool to know what these structures do}'', in conjunction with an app. A participant commented: ``\textit{It is a very simple tool to explain where things are. For example, I broke my metatarsal and could not tell much from the X-rays. With this, I would have understood}''.
}

\vspace{-5pt}

\section{Discussion}
\label{sec:discuss}

Concerning our requirements, we evaluated the sliceforms with people without visualization and/or anatomical knowledge, and we found that they could all complete their sliceforms, and saw a value in them \textbf{(R1)}. 
Our workflow is able to tackle both meshes and volumes \textbf{(R2)}, as shown in the examples of Sections~\ref{sec:mesh_results} and~\ref{sec:vol_results}.
We did not test the interaction of the user with the workflow, but it requires only three clicks (loading the data, slicing the volume, packing), which could also be reduced to one (loading the data) \textbf{(R3)}.
The outcome of the workflow is an engaging physical model, as resulting from the UE part of our evaluation, although it is not always easy to assemble \textbf{(R4)}. 
This, however, relates more to the personal aptitude of each person. 
Some evaluation participants found the assembly easier than others, but most of them learned quickly how to assemble the sliceforms, and became faster. 
The assembly of the physical model was intuitive, when people were using the order of the assembly and the instructions, as we discussed in the UP and CV part of the evaluation. 
All participants could complete their sliceforms in reasonable times, and the cost of one sliceform is less than~\EUR1, while printing requires a home printer \textbf{(R5, R6)}.
\XX{An initial evaluation was conducted to assess sliceform in a learning context, but a more thorough (and potentially comparative) study with laymen has to be performed in the future, where the sliceforms can be tested against traditional methods of anatomical education, such as anatomical illustrations from textbooks, or on-screen 3D illustrations, and against anatomical education settings in VR~\cite{preim2018survey}.}

Regarding the material, using transparent films provides a more insightful pseudo-3D view on the data than normal paper. 
With more slices, the view might become overloaded, especially if many different structures with different color assignments are present.
The volume datasets look aesthetically more pleasing, but this could be due to the fact that the mesh datasets consist of many different structures, with different visual properties. 
In this work, the selection and fine-tuning of the employed transfer function was considered to be out of scope.
However, \XX{the selection of visual properties, the order of slices, and the superimposition of slices within the physical model} plays a very big role in the \XX{visual} outcome of the sliceforms.
In the future, it should be more thoroughly investigated.
\XX{This implies that transfer functions are not only a challenging research topic for on-screen visualizations~\cite{ljung2016state}---but also, for physical visualizations.}
Additional constraints on the octree level-setting could prioritize user-specified structures.
The size of the sliceforms, and the slice distances also influence the complexity of assembly and visibility.
Small structures, such as blood vessels, might not be so visible in small or less dense constructs.
The background also affects visibility, and a white background should be preferred.
In the assembly order optimization, more sophisticated constraints could control the complexity of the papercraft, in addition to the computation of permutation performance.
For the paper placement, we are now supporting only rectangular shapes, but sliceforms could ``follow'' the geometry of the depicted data, for a more organic appearance.
Moreover, in the current approach, we allow users to define the camera position manually. An automatic approach, such as mutual information~\cite{Ohtaka:2017:SG}, could be introduced, to find the best viewpoint, before slicing.
\vspace{-6pt}

\section{Conclusions and Future Work}
\label{sec:conclude}

We present the first computer-generated crafting approach for anatomical edutainment, which uses accessible technologies (common printers) and inexpensive materials (paper or semi-transparent films) to generate cost-effective assemblable data sculptures (semi-transparent sliceforms), from anatomical meshes or volumetric medical imaging data. 
We introduce an octree formulation, and optimization for sliceform papercrafts, to support volume data based on a given transfer function, and we employ a packing that preserves the optimal assembly order and the structure of the volume, while minimizing the used resources.
Among directions for future work, we prioritize the integration of slices that ``follow’’ the geometry of the anatomical structures, where slices are not necessarily rectangular or equidistant, as well as the ability to fine-tune the visual properties of the data in a more flexible way. 
Visual indicators could also be integrated to make the assembly easier and more intuitive, while an additional evaluation to assess further the educational purpose of the sliceforms (not only the engagement and feasibility) would be required---also, compared to virtual applications. 
All in all, \textit{Slice and Dice} is an initial positive step towards interactive and engaging physicalizations for anatomical edutainment.
\vspace{-5pt}

\bibliographystyle{eg-alpha-doi}  
\bibliography{paper}        



\end{document}